\documentclass[twocolumn]{aastex62}
\usepackage{txfonts}
\usepackage[dvipdfmx]{bxpdfver}

\usepackage{color,bm}

\newcommand{\eqref}[1]{(\ref{#1})}

\shorttitle{Clouds of Fluffy Aggregates}
\shortauthors{Ohno, Okuzumi, \& Tazaki}

\begin{document}

\title{Clouds of Fluffy Aggregates: How They Form in Exoplanetary Atmospheres and Influence Transmission Spectra}

\author{Kazumasa Ohno}
\affil{Department of Earth and Planetary Sciences, Tokyo Institute of Technology, Meguro, Tokyo, 152-8551, Japan}

\author{Satoshi Okuzumi}
\affil{Department of Earth and Planetary Sciences, Tokyo Institute of Technology, Meguro, Tokyo, 152-8551, Japan}

\author{Ryo Tazaki}
\affil{Astronomical Institute, Tohoku University, 6-3, Aramaki, Aoba-ku, Sendai, Miyagi, 980-8578, Japan}

\begin{abstract}
Transmission spectrum surveys have suggested the ubiquity of high-altitude clouds in exoplanetary atmospheres.
Theoretical studies have investigated the formation processes of the high-altitude clouds; however, cloud particles have been commonly approximated as compact spheres, which is not always true for solid mineral particles that likely constitute exoplanetary clouds.
Here, we investigate how the porosity of cloud particles evolve in exoplanetary atmospheres and influence the cloud vertical profiles.
We first construct a porosity evolution model that takes into account the fractal aggregation and the compression of cloud particle aggregates.
%The compression is responsible to large aggregates, and 
Using a cloud microphysical model coupled with the porosity model, we demonstrate that the particle internal density can significantly decrease during the cloud formation.
As a result, fluffy-aggregate clouds ascend to altitude much higher than that for compact-sphere clouds assumed so far.
%The compression due to the gas drag occurs once the aggregate size exceeds $\sim {30}~{\rm \mu m}$, but it hardly occurs because the aggregates rarely grow into such large size.
We also examine how the fluffy-aggregate clouds affect transmission spectra.
We find that the clouds largely obscure the molecular features and produce a spectral slope originated by the scattering properties of aggregates.
Finally, we compare the synthetic spectra with the observations of GJ1214 b and find that its flat spectrum could be explained if the atmospheric metallicity is sufficiently high (${\ge}100\times$ solar) and the monomer size is sufficiently small ($r_{\rm mon}{<}1~{\rm \mu m}$).
The high-metallicity atmosphere may offer the clues to explore the gas accretion processes onto past GJ1214b.
\end{abstract}


%%%%%%%%%%%%%%%%%%%%%
%%%%%%%%%%%%%%%%%%%%%
\section{Introduction} \label{sec:intro}
{Transmission spectroscopy} is {a powerful approach} to probe the compositions of exoplanetary atmospheres 
\citep[e.g.,][]{Seager&Sasselov00,Brown01}. 
Recent surveys of transmission spectra have {shown} that clouds and/or hazes are {ubiquitous} in exoplanetary atmospheres \citep[e.g.,][]{Bean+10,Narita+13a,Narita+13b,Kreidberg+14,Kreidberg+18,Knutson+14a,Knutson+14b,Sing+16,Crossfield&Kreidberg17,Lothringer+18,Espinoza+19,Benneke+19}.
{A remarkable feature of the exoplanet clouds/hazes is that {some of them} are present at extremely high altitude.}
For example, the Neptune-sized exoplanet GJ436b and super-Earth GJ1214b are suggested to have an opaque cloud/haze at an altitude as high as $\sim 0.01$--$1~{\rm mbar}$ \citep{Knutson+14a,Kreidberg+14}.
The presence of the {high-altitude clouds/hazes are also suggested} for many hot Jupiters \citep[e.g.,][]{Sing+16,Barstow+17}.
{Understanding how the high-altitude clouds/hazes form may enable us to infer what composition the atmosphere beneath the clouds would have, which in turn might tell us how the planets formed.}

%{Theoretically, it is challenging to explain the presence of the the high-altitude clouds/hazes.}
In {hot, close-in} transiting planets, clouds made of condensed minerals potentially form \citep[][]{Morley+12}, and several studies have investigated their formation processes using 1D cloud microphysical models \citep[e.g.,][]{Helling+08,Helling+17,Helling+19,Lee+15,Ohno&Okuzumi18,Powell+18,Gao&Benneke18,Ormel&Min19} as well as 3D models \citep[e.g.,][]{Lee+16,Lines+18,Lines+19,Roman&Rauscher19}.
Nevertheless, it is still highly uncertain how the high-altitude clouds are formed.
\citet{Morley+13,Morley+15} and \citet{Charnay+15,Charnay+15b} showed that a high-altitude cloud producing the flat transmission spectrum of GJ1214 b could be formed if the sedimentation velocity of the particles constituting the cloud is sufficiently slow.
%\citep[for GJ436b, see][]{Morley+17}.
%\citet{Charnay+15,Charnay+15b} showed that a mineral cloud can explain the flat transmission spectrum of GJ1214 b if the sedimentation velocity of the particles constituting the cloud is sufficiently slow.
%\citep[for GJ436b, see][]{Morley+17}.
Recently, \citet[][]{Ohno&Okuzumi18} modeled the formation of clouds in GJ1214 b and GJ436 by {explicitly} calculating the size and settling velocity of cloud particles from the microphysics of particle growth.
They found that the cloud particles grow too large to ascend to a height of $0.01~\rm mbar$, needed to explain the transmission spectrum of GJ1214 b.
%as suggested from the transmission spectrum of GJ1214 b.
\citet{Gao&Benneke18} also attempted to reproduce the cloud structure of GJ1214 b using a microphysical model that fully solves the evolution of size distribution.
However, they concluded the high-altitude cloud of GJ1214 b can only be explained when the eddy diffusion coefficient in the atmospheres is assumed to be at least an order of magnitude higher than {predicted from the general circulation model (GCM) with passive tracers} \citep[][]{Charnay+15}. 
%\citet{Powell+18} showed that CARMA does not reproduce the Reyleigh scattering slope observed for many hot Jupiters \citep{Sing+16,Barstow+17}.
Photochemical hazes may explain the observed spectra if the haze production rate is  high enough \citep{Morley+15,Kawashima&Ikoma18,Adams+19,Lavvas+19}. %\citep{Morley+15,Gao+17b,Lavvas&Koskinen17,Kawashima&Ikoma18,Moran+18,Adams+19,Kawashima+19}. 
However, the haze production rate in exoplanetary atmospheres is still highly uncertain, and further laboratory studies \citep[e.g.,][]{Horst+18,He+18c,He+18b} are needed to {draw} robust conclusions.

%{\it Insert a topic sentence. It may begin with ``In this study, we propose that ...``'' or like that.}
In this study, we propose that the high-altitude cloud might be the consequence of the porosity evolution of cloud particles.
Previous studies commonly assumed that the cloud particle is a compact sphere, but this is not always true for solid condensate particles, as known from the presence of snowflakes in terrestrial atmosphere.
Theoretical and experimental studies have suggested that solid particles grow into fluffy aggregates with very low internal density \citep[e.g.,][]{Dominik&Tielens97,Blum&Wurm00,Wada+08}.
Because the fluffy aggregate has a sedimentation velocity much lower than the compact sphere with same mass, it would easily ascend to the high altitude.
Some previous studies pointed out the importance of particle porosity on the vertical structures of mineral clouds \citep[][]{Marley+13,Ohno&Okuzumi18}.
The effect of porosity evolution was recently studied for photochemical haze formation \citep{Adams+19,Lavvas+19}, while quantitative investigations have not been carried out yet for mineral cloud formation.
%however, the quantitative investigations have not been carried out yet.

In this paper, we investigate how the porosity of cloud particles evolve in exoplanetary atmospheres, and how it affects the vertical profiles of mineral clouds. 
Using a cloud microphysical model coupled with the porosity evolution model, we will demonstrate that cloud particle aggregates (CPAs, hereafter) grow without compression in most cases studied here.
% using a microphysical model that takes into account the vertical transport, particle growth, and porosity evolution in a self-consistent manner.
%\textcolor{blue}{We will demonstrate that the compression of cloud particle aggregates can be negligible in most cases studied here.}
We will also {compute synthetic transmission spectra to study} how the fluffy-aggregate clouds influence the observable transmission spectra.
%observational signature{s} of the fluffy-aggregate clouds.
%\textcolor{blue}{Our study examines how the porosity of aggregates evolves using the physically-based model and thus complements a recent study of \citet{Adams+19}, who investigated the effects of the porosity evolution on haze formation using the empirical model.}
The organization of this paper is as follows.
We introduce how internal density of an aggregate varies with microphysical processes and establish a porosity evolution model for CPAs in Section \ref{sec:method}.
We describe our microphsical model and investigate the vertical structures of the fluffy-aggregate clouds in Section \ref{sec:vertical}.
We present the synthetic transmission spectrum and compare it with the observations of GJ1214 b in Section \ref{sec:result2}.
We discuss the caveats of this study and future prospects in Section \ref{sec:discussion}.
We summarize this paper in Section \ref{sec:summary}

%%%%%%%%%%%%%%%%%%%%%%%%%%
%\section{Model}\label{sec:method}
%%%%%%%%%%%%%%%%%%%%%%%%%%%%%%%%%%%%%%%%%%%%%%%%%%%%%%%%%%%%%%%%%%%%%%%%%%%%%%%
\section{Modeling the Formation of Fluffy Aggregates in Mineral Clouds}\label{sec:method}
%%%%%%%%%%%%%%%%%%%%%%%%%%%%%%%%%%%%%%%%%%%%%%%%%%%%%%%%%%%%%%%%%%%%%%%%%%%%%%%

{Fluffy aggregates form} through the mutual sticking of solid particles with a low collision energy \citep[e.g.,][]{Meakin91}. 
The smallest particles constituting an aggregate are called {\it monomers}.
%%%%%%%%%%%%%%%%%%%%%%%%%%%%%%%%%%%%%%%%%%%%%%%%%%%%%%%%%%%%%%%%%%%%%%%%%%%%%%%
%\subsection{Quantities Characterizing Fluffy Aggregates}
%%%%%%%%%%%%%%%%%%%%%%%%%%%%%%%%%%%%%%%%%%%%%%%%%%%%%%%%%%%%%%%%%%%%%%%%%%%%%%%
One of the most important quantities that characterize a porous aggregate is the filling factor $\phi$ defined by
\begin{equation}\label{eq:internal-evolution}
\phi = \frac{\rho_{\rm agg}}{\rho_{\rm mon}},
\end{equation}
where $\rho_{\rm agg}$ is the mean internal density of the aggregate and $\rho_{\rm mon}$
are the bulk density of the individual monomers.
For aggregates made of single-sized monomers, 
Equation~\eqref{eq:internal-evolution}
can also be written as
\begin{equation}
\label{eq:phi_def}
\phi = \frac{NV_{\rm mon}}{V_{\rm agg}},
\end{equation}
where $N$ is the number of the constituent monomers,
and $V_{\rm agg}$ and $V_{\rm mon}$ are the volumes of the aggregate and individual monomers, respectively. 
Here, the volume of an aggregate is defined as that of a sphere with the same gyration radius.
The number of constituent monomers is another important parameter for aggregate of monodisperse monomers because it is directly related to the aggregate mass.

The set of $N$ and $\phi$ defines the characteristic size, or length scale, of a porous aggregate.
If we approximate an aggregate with a sphere of radius $r_{\rm agg}$, the ratio of volumes $V_{\rm mon}$ to $V_{\rm agg}$ is $V_{\rm agg}/V_{\rm mon} = (r_{\rm agg}/r_{\rm mon})^3$, where $r_{\rm mon}$ is the monomer radius. 
%If we approximate an aggregate with a sphere of radius $r_{\rm agg}$, the volumes $V_{\rm agg}$ and $V_{\rm mon}$ are $r_{\rm agg}$ and the monomer radius $r_{\rm mon}$ as $V_{\rm agg}/V_{\rm mon} = (r_{\rm agg}/r_{\rm mon})^3$. 
Using this expression with Equation~\eqref{eq:phi_def},   
we obtain the relation that determines 
$r_{\rm agg}$ as a function of $N$ and $\phi$,
%{generally} have {the} following relation,
\begin{equation}
\label{eq:mass-relation}
    % \phi =\left(\frac{r_{\rm mon}}{r_{\rm agg}}\right)^{3}N.
    r_{\rm agg} = \left(\frac{N}{\phi}\right)^{1/3}r_{\rm mon},
\end{equation}

In atmospheres, the filling factor of an aggregate can change through various processes, and we introduce them in following subsections.

\subsection{Evolution of the Filling Factor}
We introduce how the filling factor of an aggregate $\phi$ evolves via various processes. For convention, we describe a filling factor determined by a specific process using a subscript $\phi$; for example, $\phi_{\rm coll}$ for the collisional compression.
\subsubsection{Fractal Growth}\label{sec:fractal}
Aggregates forming through low-energy sticking collisions often have an open structure with fractal geometry \citep[e.g.,][]{Meakin91}. A fractal aggregate can be characterized by 
the fractal dimension $D_{\rm f}$ defined by 
\begin{equation}\label{eq:Df}
N=k_{\rm 0}\left( \frac{r_{\rm agg}}{r_{\rm mon}}\right)^{D_{\rm f}},
\end{equation}
where $k_{\rm 0}$ is {a prefactor of order unity}, $r_{\rm mon}$ is the radius of {individual monomers}, and $r_{\rm agg}$ is the characteristic radius of an aggregate. 
An aggregate with $D_{\rm f}=1$ {is ``chain-like'' in the sense that its length scale $r_{\rm agg}$ is proportional to its mass ($\propto N$), while an aggregate with $D_{\rm f}=2$ is ``plane-like'' in the sense that its cross section $\sim r_{\rm agg}^2$ is proportional to its mass.} %{\it Note: $D_{\rm f}=3$ does not necessarily mean a sphere}. 
Experimental and numerical studies {show} that {aggregates growing by accreting similar-sized aggregates have $D_{\rm f}=1.7$--$2.2$}, whereas aggregates growing by accreting individual monomers tend to have $D_{\rm f}\approx3$ \citep[e.g.,][]{Meakin91,Okuzumi+09}.
Non-ballistic collisions and rotation of aggregates could also reduce the fractal dimension down to $D_{\rm f}\approx 1.1$ \citep{Paszun&Dominik06}.
$D_{\rm f}=2$ is often assumed in the studies of haze formation on Titan and Pluto \citep[e.g.,][]{Lavvas+10,Gao+17a}.
{Unless otherwise noted, we assume that aggregate-aggregate collisions dominate over aggregate-monomer collisions, adopting $D_{\rm f}\approx 2$ and $k_0 \approx 1$ \citep{Okuzumi+09}.}
%The assumption has been often adopted in the studies of haze formation on solar system objects \citep[e.g.,][]{Lavvas+10,Gao+17a}.
We will discuss the validity of the assumption in Section \ref{sec:discussion}.
%It is also suggested that the observations of the photochemical haze in Titan and Pluto are reasonably explained by fractal aggregates with $D_{\rm f}=2$ \citep[e.g.,][]{Tomasko+08,Gao+17a}.

Once the fractal dimension is given, the filling factor of a fractal aggregate, $\phi_{\rm frac}$, can be calculated as a function of $N$.
Substituting $r_{\rm agg}/r_{\rm mon} = k_0^{-1/D_{\rm f}}N^{1/D_{\rm f}}$ along with $D_{\rm f}=2$ and $k_0 = 1$ into {Equation} \eqref{eq:mass-relation} and solving for $\phi$, we obtain 
% the filling factor determined by the fractal growth is given by
% \begin{equation}\label{eq:phi_frac}f
%     \phi_{\rm frac}=k_{\rm 0}\left( \frac{r_{\rm agg}}{r_{\rm mon}}\right)^{D_{\rm f}-3},
% \end{equation}
% or given as a function of the number of monomers,
\begin{equation}
    % \phi_{\rm frac}=k_{\rm 0}^{3/D_{\rm f}}N^{1-3/D_{\rm f}}.
    \phi_{\rm frac}= N^{-1/2},
    \label{eq:phi_frac}
\end{equation}
which indicates that the filling factor decreases with increasing $N$, i.e., as the aggregate grows. Whenever two aggregates stick at a low velocity, the newly formed aggregate contains a large void whose volume is comparable to the volume of the collided aggregates (see Section 4 of \citealt{Okuzumi+09} for more quantitative analysis). This causes the decrease of the filling factor.

\subsubsection{Collisional Compression} \label{sec:phi_coll}
The fractal growth described by Equation~\eqref{eq:phi_frac} breaks down if the impact energy is higher than needed for internal restructuring of the newly forming aggregate, for which case collisional compaction occurs \citep[e.g.,][]{Dominik&Tielens97,Blum&Wurm00,Wada+07,Wada+08,Paszun&Dominik09}.
{For a collision between two aggregates with similar individual masses $\approx m_{\rm agg}/2$,} the collisional energy can approximately be written as 
\begin{equation}\label{eq:E_imp}
    E_{\rm imp} \approx \frac{1}{8}m_{\rm agg}{\Delta v}^2,
\end{equation}
where $\Delta v$ is the collisional velocity. {Here, $m_{\rm agg}$ stands for the mass of the newly forming aggregate, and we have used that the reduced mass of the collided aggregates is $\approx (m_{\rm agg}/2)/2 = m_{\rm agg}/4$.}
Restructuring of the new aggregate occurs if $E_{\rm imp}$ is much higher than the energy $E_{\rm roll}$ needed to roll one monomer over another monomer {in contact} by ${90}^{\rm \circ}$ against rolling friction \citep{Dominik&Tielens97,Blum&Wurm00}.
Following \citet{Dominik&Tielens95}, we evaluate $E_{\rm roll}$ as
\begin{equation}\label{eq:E_roll}
E_{\rm roll}=6\pi^2 \gamma r_{\rm mon}\xi_{\rm crit},
%&\approx&6\times{10}^{-16}~{\rm J}~\left( \frac{\gamma}{0.1~{\rm J~m^{-2}}}\right)\left( \frac{r_{\rm mon}}{1~{\rm \mu m}}\right)\left( \frac{\xi_{\rm crit}}{1~\AA}\right),
\end{equation}
where $\gamma$ is the surface energy of the monomers and $\xi_{\rm crit}$ is the critical rolling displacement {above which inelastic rolling occurs.} 
A realistic value of ${\xi}_{\rm{crit}}$ is somewhat uncertain: the model of \citet{Dominik&Tielens95} anticipates ${\xi}_{\rm{crit}}{\sim}2~\AA$, whereas the measurement by \citet{Heim+99} of the rolling friction force acting on silica microspheres suggests a $\sim 10$ times larger value.
We set  ${\xi}_{\rm{crit}} = 2~{\rm \AA}$ to examine maximal impacts of the compression processes.

The filling factor of grain aggregates after collisional internal restructuring has been extensively studied by means of $N$-body dynamical simulations \citep{Wada+07,Wada+08,Paszun&Dominik09,Suyama+08,Suyama+12}.
According to \citet{Wada+08}, the size of an aggregate after a {high-energy ($E_{\rm imp} \ga E_{\rm roll}$) collision between two equal-sized fractal ($D_{\rm f}=2$) aggregates} follows
\begin{equation}\label{eq:Wada}
    \frac{r_{\rm agg}}{r_{\rm mon}}=N^{2/5}\left( \frac{E_{\rm imp}}{0.15NE_{\rm roll}}\right)^{-1/10}.
\end{equation}
%{{\it The following sentence is unnecessary and should be deleted:} Numerical simulations suggest that, even if the collisional velocity so high that maximum compression occurs ($E_{\rm imp}\sim (0.1$--$1)NE_{\rm roll}$), the aggregate still retains the fractal dimension of $D_{\rm f}\approx 2.5$ \citep{Wada+08,Suyama+08}.}
Using Equation \eqref{eq:mass-relation}, 
Equation \eqref{eq:Wada} translates into the filling factor after a high-energy collision,
\begin{equation}\label{eq:phi_coll}
    \phi_{\rm coll}=N^{-1/2}\left( \frac{E_{\rm imp}}{0.15E_{\rm roll}}\right)^{3/10}.
\end{equation}
{Here,} the prefactor $N^{-1/2}$ corresponds to the filling factor {without collisional compression (see Equation~\ref{eq:phi_frac}), whereas the factor 
$({E_{\rm imp}}/{0.15E_{\rm roll}})^{3/10}$
represents compression occurring for 
$E_{\rm imp} \ga E_{\rm roll}$. 
\citet{Wada+08} derived Equation~\eqref{eq:Wada} for aggregates after a single compressive collision, but \citet{Suyama+08} later confirmed that the expression approximately holds for aggregates growing through multiple compressive collisions (see their Equation~(33)).
}

%\textcolor{blue}{
{For particles in atmospheres, the collision velocity in Equation~\eqref{eq:E_imp}} is calculated as the root sum square of the thermal (Brownian) relative velocity and the relative velocity $\Delta v_{\rm t}$ of gravitational settling, i.e., 
\begin{equation}\label{eq:vrel}
    \Delta v {=} \sqrt{\frac{32k_{\rm B}T}{\pi m_{\rm agg}}+ \Delta v_{\rm t}^2}.
\end{equation}
%where we have used that the reduced mass of aggregates with similar mass $\approx m_{\rm agg}/2$ is $m_{\rm agg}/8$.
Here we write $\Delta v_{\rm t} \approx \epsilon v_{\rm t}'$, where $v_{\rm t}'$ is the terminal settling velocity of individual aggregates before collision and $\epsilon$ is a numerical factor arising from finite width of actual size distribution of the aggregates.
We here adopt $\epsilon = 0.5$ following \citet{Sato+16}.
For the terminal velocity of aggregates, we use an expression for spheres \citep{Ohno&Okuzumi17},
\begin{equation}\label{eq:vt}
v_{\rm t}' =\frac{2gr_{\rm agg}'^2\rho'_{\rm agg}}{9\eta}\beta(r_{\rm agg}') \left[ 1+\left( \frac{0.45g r_{\rm agg}'^3\rho_{\rm g}\rho'_{\rm agg}}{54\eta^2}\right)^{2/5}\right]^{-5/4},
\end{equation}
where $r_{\rm agg}'$ and $\rho'_{\rm agg}$ are the characteristic radius and density of aggregates before collision, respectively, $\eta$ is the dynamic viscosity of ambient gas, and $\beta$ is the slip correction factor accounting for the free-molecular flow regime. 
In Equation \eqref{eq:vt}, we have approximated the aerodynamic radius of an aggregate with its characteristic radius $r_{\rm agg}$ defined by Equation \eqref{eq:mass-relation}. This approximation is invalid for very fluffy aggregates with $D_{\rm f}<2$, for which the aerodynamic radius is generally smaller than the characteristic radius \footnote{The reason can be easily understood for the special case of the free molecular regime, for which the aerodynamic cross section is approximately equal to the projected area \citep{Blum+96}. For $D_{\rm f} < 2$, the projected area increases linearly with mass \citep[e.g.,][]{Minato+06}, but the ``characteristic'' cross section $\pi r_{\rm agg}^2 \propto N^{2/D_{\rm f}}$ increases {\it faster} than mass ($\propto N$). For $D_{\rm f} \approx 2$, the characteristic cross section $\approx N\pi r_{\rm mon}^2$ is only $\sim 2$ times larger than the projected area \citep[see, e.g., Figure 8 of][]{Okuzumi+09}, and therefore the characteristic radius differs from the  aerodynamic radius only by $\sim 40\%$. The approximation is even better for $D_{\rm f} > 2$ \citep{Okuzumi+09}.} \citep{Okuzumi09}.
We use this assumption because we only consider $D_{\rm f} \geq 2$ in this study. 
The slip correction factor is given by \citep[e.g.,][]{Seinfeld&Pandis06}
\begin{equation}
\beta(r_{\rm agg}') = 1+
\frac{l_{\rm g}}{r_{\rm agg}'}\left[1.257+0.4\exp\left(-\frac{1.1r_{\rm agg}'}{l_{\rm g}}\right)\right],
\end{equation}
where 
$l_{\rm g}$ is the mean free path of gas molecules.
The second term in the bracket in Equation \eqref{eq:vt} corrects for high Reynolds (turbulent) flow, although it is mostly negligible for slowly settling aggregates considered in this study.
%}

%%%%%%%%%%%%%%%%%%%%%%%%%%%%
\subsubsection{Gas-drag Compression}\label{sec:porosity_recipe}
An aggregate moving relative to the surrounding gas can experience compression when the gas drag force acting on it is strong enough to cause internal restructuring.  
We employ the model of \citet{Kataoka+13a} to evaluate the filling factor of an aggregate under gas-drag compression (see \citealt{Kataoka+13a,Arakawa&Nakamoto16} for applications of the model to dust evolution in protoplanetary disks).
We assume that compression occurs when the ram pressure $P_{\rm ram}$ of the gas flow exceeds the static compressional strength $P_{\rm str}$ of the aggregate. The compression thus proceeds until $P_{\rm ram}$ becomes equal to $P_{\rm str}$. Based on the results of $N$-body simulations, \citet{Kataoka+13a} found that the static compressional strength can be written as
%compression the equilibrium filling factor for the static compression is determined by the balance between the ram pressure and the compression strength $P_{\rm str}$, given by
\begin{equation}\label{eq:P_str}
P_{\rm str}=\frac{E_{\rm roll}}{r_{\rm mon}^3}\phi^3,
\end{equation}
where $E_{\rm roll}$ is the rolling energy already introduced in Section~\ref{sec:phi_coll}.
The ram pressure can be evaluated as {the drag force per cross section of the aggregate. For an aggregate setting in an atmosphere at a terminal velocity, the drag force is equal to the gravity $m_{\rm agg}g$, where $g$ is the gravitational acceleration.} Thus, $P_{\rm ram}$ is given by 
\begin{equation}\label{eq:P_gas}
P_{\rm ram}\approx \frac{m_{\rm agg}g}{\pi r_{\rm agg}^2}
=\frac{4}{3}r_{\rm agg}g\rho_{\rm mon}\phi.
\end{equation}
{Solving $P_{\rm str} = P_{\rm ram}$ together with Equation~\eqref{eq:mass-relation} for $\phi$}, the equilibrium filling factor {under gas-drag} compression is obtained as
{
\begin{equation}\label{eq:phi_drag}
\phi_{\rm drag}=N^{1/7}
\left( \frac{4g\rho_{\rm mon}r_{\rm mon}^4}{3E_{\rm roll}}\right)^{3/7}.
%\phi_{\rm drag}=N^{1/4}
%\left( \frac{4g\rho_{\rm mon}r_{\rm mon}^4}{3E_{\rm roll}}\right)^{1/2}.
\end{equation}
}
Equation \eqref{eq:phi_drag} indicates that {under gas-drag compression, the filling factor increases with aggregate mass}. 
It is worth noting that {$\phi_{\rm drag}$} is independent of {the} ambient gas density because the gas drag force {balances with the gravity, which does not depend on the gas density}. 

%%%%%%%%%%%%%%%%%%%%%%%%%%%%%%%
\subsubsection{A General Formula}
For a given number of monomers, equivalent to the aggregate mass, one can calculate the equilibrium filling factor from the highest one determined by the fractal growth, gas-drag compression, and collisional compression \citep{Kataoka+13b}, i.e., 
\begin{equation}\label{eq:phi_eq}
    \phi_{\rm eq}={\rm max}[\phi_{\rm frac},\phi_{\rm drag},\phi_{\rm coll}].
\end{equation}
%The expression enables us to evaluate the filling factor of aggregates 

%%%%%%%%%%%%%%%%%%%%%%%%%%%%%%%%%%%%
\subsection{An Example: KCl Cloud Aggregates in GJ1214b}\label{sec:kataoka}
%%%%%%%%%%%%%%%%%%%%%%%%%%%%%%%%%
\begin{figure*}[t]
\includegraphics[clip,width=\hsize]{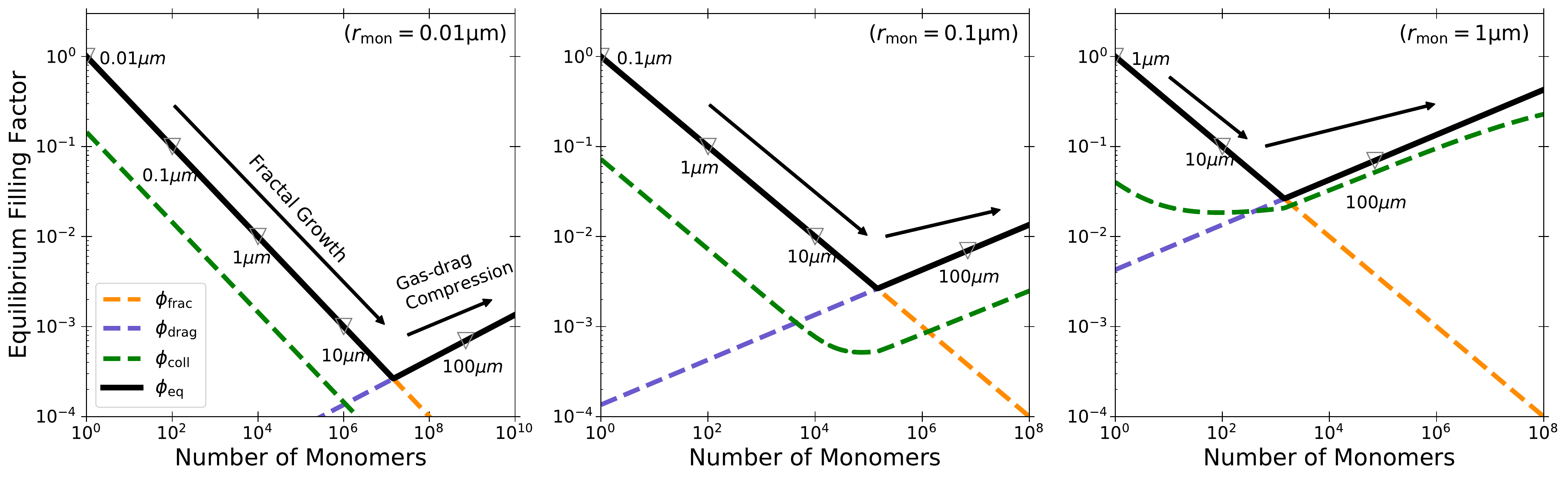}
\caption{Equilibrium filling factor of {KCl particle aggregates at the base of the KCl cloud in the super-Earth GJ1214b}. %\edit1{\it Please indicate $\phi_{\rm eq}$ by solid lines and show three contributions by thin (or dasehed/dotted) lines.} 
The left, center, and right panels are for monomer radii $r_{\rm mon}=0.01$, $0.1$, and $1~{\rm \mu m}$, respectively.
The orange, blue, green, and black lines show the filling factors determined by fractal growth ($\phi_{\rm frac}$; Equation \ref{eq:phi_frac}), gas-drag compression ($\phi_{\rm drag}$; Equation \ref{eq:phi_drag}), collisional compression ($\phi_{\rm coll}$; Equation \ref{eq:phi_coll}), and all of them ($\phi_{\rm eq}$; Equation \ref{eq:phi_eq}), respectively.
The aggregate radius $r_{\rm agg}=0.01$, $0.1$, $1$, $10$, and $100~{\rm \mu m}$ are denoted as the triangles.
 }\label{fig:kataoka}
\end{figure*}
%%%%%%%%%%%%%%%%%%%%%%%%%%%%%%%
We {here illustrate} how the filling factor of CPAs in an super-Earth atmosphere evolves as they grow.
We consider the cloud of KCl solid particles in the super-Earth GJ1214b. 
It is assumed that the cloud has its base at $P=100~{\rm mbar}$ and $T=700~{\rm K}$, 
where $P$ is the atmospheric pressure. 
% The surface gravity of GJ1214b is $g=8.93~{\rm m~s^{-2}}$.  \edit1{$\gamma=0.1~{\rm J~m^{-2}}$ \it Why not using Eq. 8?}, and $\rho_{\rm mon}=2~{\rm g~{cm}^{-3}}$, corresponding to KCl cloud particles in GJ1214b.
The material density and surface energy are $\rho_{\rm mon}=2~{\rm g~{cm}^{-3}}$ and $\gamma=0.11~{\rm J~m^{-2}}$ for KCl crystals \citep{Westwood&Hitch63}.
We note that one cannot calculate the filling factor for collisional compression $\phi_{\rm coll}$ without a knowledge of filling factor of the aggregates before the collision, as the terminal velocity depends on the aggregate density (see Equation \ref{eq:vt}). 
Thus, we first calculate $\phi_{\rm eq}$ only from $\phi_{\rm frac}$ and $\phi_{\rm gas}$, and then $\phi_{\rm coll}$ is calculated with the obtained $\phi_{\rm eq}$. 
%The method enables us to study which gas-drag and collisional compression is dominant.

We find that the internal density of CPAs can be lower than the material density by several orders of magnitude.
The evolution pathways of the equilibrium filling factor for $r_{\rm mon} = 0.01$, 0.1, and 1 $\micron$ are shown in Figure \ref{fig:kataoka}. 
Here the equilibrium filling factor is expressed as a function of the number of monomers making up the aggregates, $N = m_{\rm agg}/m_{\rm mon}$. 
One can see that the aggregates are highly porous, with $\phi_{\rm eq} \la 0.1$, over a wide range of $N$. For small $N$, both gas-drag and collisional compression are negligible and the filling factor is determined by fractal growth.
Once an aggregate size exceeds a certain value, either collisional or gas-drag compression sets in.
%and the filling factor decreases as an aggregate grows into large size.
%Once an aggregate size exceeds a certain value, the aggregate starts to experience compression.
For all monomer sizes shown in Figure \ref{fig:kataoka} ($r_{\rm mon}=0.01$--$1~{\rm \mu m}$), gas-drag compression always dominates over collision compression.
Collisional compression is important for larger monomer sizes and occurs only for $r_{\rm mon} \ga 1~{\rm \mu m}$ around at the cloud base.
No matter which compression mechanism dominates, the filling factor increases with $N$, and hence with aggregates mass. 
Nevertheless, the filling factor never exceeds 0.1 as long as the monomer mass is in the range $10^2 \la N \la 10^6$. The results thus demonstrate the importance of considering the porosity of mineral cloud aggregates. 

%%%%%%%%%%%%%%%%%%%%%%%%%%%
\subsection{Analytic Estimates of Compression Threshold Sizes}
%%%%%%%%%%%%%%%%%%%%%%%%%%%
To further elaborate how the porosity of CPAs evolve in general cases, we here analytically estimate the threshold sizes at which the compression sets in. 
%To achieve a general understanding of how the porosity of cloud particle aggregates evolve in general cases, we here analytically estimate the threshold sizes at which compression processes become effective. 
%One can evaluate the filling factor of aggregates from the fractal growth for sizes smaller than the threshold.

%%%%%%%%%%%%%%%%%%%%%%%%%%%%%%%%%%
\subsubsection{Gas-drag Compression Threshold}
%%%%%%%%%%%%%%%%%%%%%%%%%%%%%%%%%
% quantitatively argue criteria for occuring the compression, which will be useful to quantify the impacts on actual cloud formation.
Comparison between Equations \eqref{eq:phi_frac} and \eqref{eq:phi_drag} shows that $\phi_{\rm drag}$ exceeds $\phi_{\rm frac}$ when the number of monomers satisfies
\begin{equation}
    N>\left( \frac{9\pi^2 \gamma \xi_{\rm crit}}{2\rho_{\rm mon}g}\right)^{2/3}r_{\rm mon}^{-2},
\end{equation}
where we use Equation \eqref{eq:E_roll}.
Since $r_{\rm agg} = N^{1/2}r_{\rm mon}$ for $D_{\rm f} = 2$, we find that a $D_{\rm f} = 2$ aggregate starts to experience gas-drag compression when its characteristic radius exceeds a threshold 
\begin{eqnarray}\label{eq:r_comp}
r_{\rm drag}&=&\left(\frac{9\pi^2\gamma \xi_{\rm crit}}{2\rho_{\rm mon}g}\right)^{1/3} \nonumber \\
&\approx& 30~{\rm \mu m} \left( \frac{g}{10~{\rm m~s^{-2}}}\right)^{-1/3}\left( \frac{\rho_{\rm mon}}{2~{\rm g~{cm}^{-3}}}\right)^{-1/3}\left( \frac{\gamma}{0.1~{\rm J~m^{-2}}}\right)^{1/3}.
\end{eqnarray}
It is worth noting that $r_{\rm drag}$ is independent of the monomer size and only depends on material properties and planetary gravity.
Equation \eqref{eq:r_comp} indicates that gas-drag compression is responsible to aggregates larger than tens micron, while it will be responsible to micron-sized aggregates on high-gravity objects, such as brown dwarfs.
%Due to the weak dependencies, the gas-drag compression sets in at $r_{\rm agg} \sim 30~{\rm \mu m}$ for planet-sized objects.
%For objects with high surface gravity, such as brown dwarfs ($g\sim{10}^3~{\rm m~s^{-2}}$), gas-drag compression will be responsible to micron-sized aggregates.

%%%%%%%%%%%%%%%%%%%%%%%%%%%%%%%%%%
\subsubsection{Collisional Compression Threshold}
%%%%%%%%%%%%%%%%%%%%%%%%%%%%%%%%%
% In principle, the collisional compression never occurs for a small number of monomers because the filling factor with collisional compression $\phi_{\rm coll}$ is always an order of magnitude lower than the filling factor with the fractal growth $\phi_{\rm frac}$.
% This is due to the fact that collisional velocity is dominated 
We here estimate the threshold size at which fractal aggregates begin to be compressed by high-energy collisions.  
Since the thermal kinetic energy $k_{\rm B}T{\sim}{10}^{-20}~{\rm J}~(T/1000~{\rm K})$ is generally several orders of magnitude smaller than the rolling energy $E_{\rm roll}{\sim}{10}^{-17}~{\rm J}~(\gamma/0.1~{\rm J~m^{-2}})(r_{\rm mon}/1~{\rm \micron})$, one can consider that only relative velocity from gravitational settling induces collisional compression. % On the other hand, relative motion caused by gravitational settling dominates the collisional energy as an aggregate grows.
For small fractal aggregates, the second term in the bracket in Equation~\eqref{eq:vt} is negligible, and thus we approximately have  
% If we crudely assume $A\approx \pi r_{\rm agg}'^2$ and the laminar flow limit in Equation \eqref{eq:vt}, terminal velocity of an uncompressed aggregate can be approximated as
\begin{equation}\label{eq:vt_appro}
    v_{\rm t}'\approx \frac{2g r_{\rm agg}'^2\rho_{\rm agg}'}{9\eta}\beta.
\end{equation}
For fractal aggregates of $D_{\rm f} = 2$, we also have $\rho_{\rm agg}' \approx (r_{\rm mon}/r_{\rm agg}')\rho_{\rm mon}$, $r_{\rm agg}' = 2^{-1/2} r_{\rm agg}$, 
and $m_{\rm agg} = (r_{\rm agg}/r_{\rm mon}
)^2 m_{\rm mon}$, where $r_{\rm agg}$ is the radius of the newly formed aggregate.
Substituting $\Delta v_{\rm t}\approx\epsilon v_{\rm t}'$ with these expressions into Equation \eqref{eq:E_imp}, the collisional energy of a settling-induced collision is given by 
\begin{equation}\label{eq:E_imp_grav}
    E_{\rm imp}\approx \frac{1}{16} m_{\rm mon}\left( 
    \frac{g r_{\rm agg}^2 \rho_{\rm mon}}{9\eta}\beta
    \right)^2. 
\end{equation}
% Equation \eqref{eq:E_imp_grav} indicates that the collisional energy increases with increasing aggregate size, and thus the collisional compression occurs when an aggregate grows into a sufficiently large size.
%The collisional compression dominates over the gas-drag compression when the collsision energy is higher than the rolling energy at the size of $r_{\rm agg}=r_{\rm drag}$.
%According to Equation \eqref{eq:phi_frac}, the collisional energy for aggregates with the same size is higher for larger monomer size.
%This is why the collisional compression occurs only for a large monomer size.
%However, as shown in later sections, growth of an aggregate constituted by large monomers is relatively inefficient, and the collisional compression hardly occurs.

Collisional compression occurs ($\phi_{\rm coll} > \phi_{\rm frac}$) when $E_{\rm imp} > 0.15 E_{\rm roll}$ (see Equations~\eqref{eq:phi_frac} and \eqref{eq:phi_coll}). 
For $r_{\rm agg} \gg l_{\rm g}$ ($\beta\approx1$), the threshold size for collisional compression is given by 
\begin{eqnarray}
    r_{\rm coll}&=&\left( \frac{9\eta}{\rho_{\rm mon}g}\sqrt{\frac{2.4E_{\rm roll}}{m_{\rm mon}}}\right)^{1/2}
    \nonumber \\
    &\approx& 70~{\rm \mu m} \left( \frac{g}{10~{\rm m~s^{-2}}}\right)^{-1/2}\left( \frac{\rho_{\rm mon}}{2~{\rm g~{cm}^{-3}}}\right)^{-3/4} \left( \frac{r_{\rm mon}}{1~{\rm \mu m}}\right)^{-1/2}\\
    \nonumber
    &&\times \left( \frac{\gamma}{0.1~{\rm J~m^{-2}}}\right)^{1/4}\left( \frac{T}{1000~{\rm K}}\right)^{1/4},
\end{eqnarray}
where we have used the dynamic viscosity for hydrogen-rich atmospheres $\eta=5.877\times {10}^{-7}~{\rm Pa~s}\sqrt{T{\rm [K]}}$ \citep{Woitke&Helling03}.
% Using $\eta=\rho_{\rm g}v_{\rm th}l_{\rm g}/3$, where $v_{\rm th}=\sqrt{8k_{\rm B}T/\pi m_{\rm g}}$ is the mean thermal velocity of gas molecules and $m_{\rm g}$ is the mass of a gas molecule, the critical size $r_{\rm eps}$ in the free molecular flow regime ($\beta\approx 1.5l_{\rm g}/r_{\rm agg}$) is also given by
In the opposite limit of $r_{\rm agg} \ll l_{\rm g}$, for which $\beta \approx 1.7 l_{\rm g}/r_{\rm agg}' \approx 2.4l_{\rm g}/r_{\rm agg}$, we obtain the threshold size of %\edit1{\it Please recalculate.}
\begin{eqnarray}\label{eq:r_coll}
    r_{\rm coll}&=& \frac{10 P}{\pi \rho_{\rm mon}gv_{\rm th}}\sqrt{\frac{2.4E_{\rm roll}}{m_{\rm mon}}}\\
    \nonumber
    &\approx& 90~{\rm \mu m} \left( \frac{g}{10~{\rm m~s^{-2}}}\right)^{-1}\left( \frac{\rho_{\rm mon}}{2~{\rm g~{cm}^{-3}}}\right)^{-3/2} \left( \frac{r_{\rm mon}}{1~{\rm \mu m}}\right)^{-1}\\
    \nonumber
    &&\times \left( \frac{\gamma}{0.1~{\rm J~m^{-2}}}\right)^{1/2}\left( \frac{v_{\rm th}}{1~{\rm km~s^{-1}}}\right)^{-1}\left( \frac{P}{100~{\rm mbar}}\right).
\end{eqnarray}
Here, we have used $l_{\rm g} = 3\eta/(\rho_{\rm g}v_{\rm th})$ and $\rho_{\rm g}=(8/\pi)P/v_{\rm th}^2$, where $v_{\rm th}=\sqrt{8k_{\rm B}T/\pi m_{\rm g}}$ is the mean thermal velocity of gas molecules and $m_{\rm g}$ is the mass of a gas molecule.
% Therefore, the threshold size for collisional compression for general cases can be approximately evaluated as
% \begin{equation}
%     r_{\rm coll}=\min[r_{\rm stk},r_{\rm eps}].
% \end{equation}
% %Since the terminal velocity for the molecular flow regime is 

%%%%%%%%%%%%%%%%%%%%%%%%%%%%%%%%%%%%%%%%%%%%%%%%%%%%%%%
\section{Vertical Structure of Fluffy-Aggregate Clouds}
\label{sec:vertical}
%%%%%%%%%%%%%%%%%%%%%%%%%%%%%%%%%%%%%%%%%%%%%%%%%%%%%%%

%%%%%%%%%%%%%%%%%%%%%%%%%%%%
\subsection{Model}\label{sec:model}
%%%%%%%%%%%%%%%%%%%%%%%%%%
%In this section, we introduce a cloud model used in this study.
To demonstrate how the porosity evolution affects the cloud structures, we calculate the vertical transport and growth of cloud particles using the double-moment bulk scheme described by \citet{Ohno&Okuzumi18}.
The model adopts a 1D Eulerian framework and calculates the vertical distributions of the mass density ($\rho_{\rm c}$) and number density ($n_{\rm c}$) of the cloud particles.
% The approach is called the double-moment bulk scheme in meteorology \citep[e.g.,][]{Ziegler85,Ferrier94}.
The model assumes that the mass distribution of particles is narrowly peaked at the characteristic mass $m_{\rm agg}$ dominating the total cloud mass.
%The narrowly-peaked mass distribution can be seen in the results of recent studies investigating the size distribution in mineral clouds with the spectral bin-scheme \citep[for example, see Figure 4 of][]{Powell+18}, although the collisional growth was not taken into account in their simulations.
In this context, the mass and number densities are related as $\rho_{\rm c}=m_{\rm agg}n_{\rm c}$.
% This approach is equivalent to the characteristic size method used to simulate grain growth in protoplanetary disks \citep[e.g.,][]{Okuzumi+11,Birnstiel+12,Sato+16}. 
%We introduce the treatment of vertical transport and growth of cloud particles in following subsections.

%%%%%%%%%%%%%%%%%%%%%%%%%%%%%%%%%%%%%%%
\subsubsection{Prescription of Nucleation and Condensation}\label{sec:nucleation}
%%%%%%%%%%%%%%%%%%%%%%%%%%%%%%%%%%%%%%%
%\edit1{\it Please try to summarize this subsubsection into two not-so-long paragraphs (e.g., one for nucleation, one for condensation). Be concise!}

\begin{figure}[t]
\includegraphics[clip,width=\hsize]{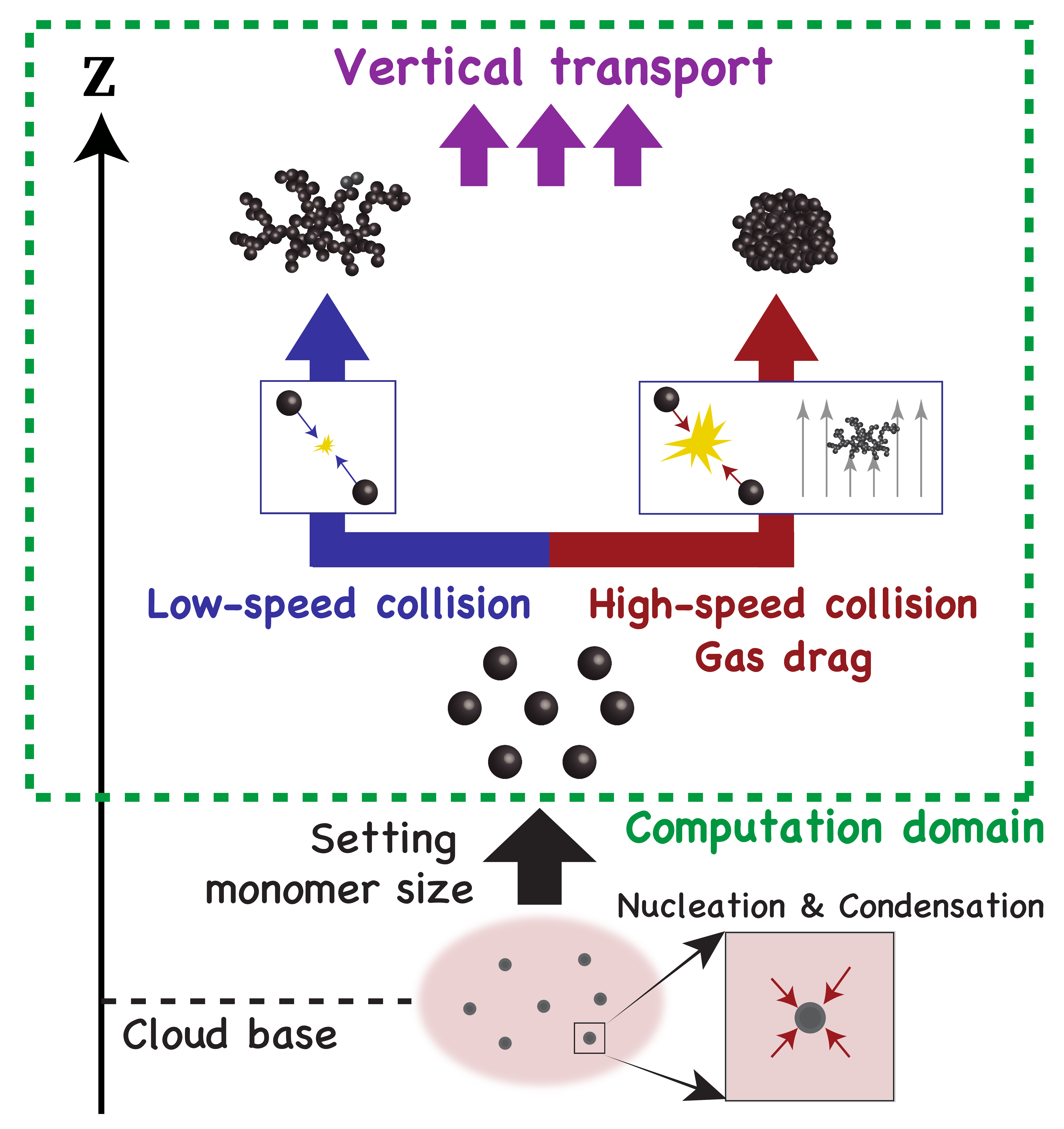}
\caption{Cartoon illustrating the formation of fluffy-aggregate clouds.}\label{fig:schematic}
\end{figure}

Formation of the fluffy-aggregate cloud will be triggered by the formation of monomers via nucleation followed by condensation (Figure \ref{fig:schematic}). The processes will determine the size of monomers, which predominantly control the porosity evolution and thus the particle growth. However, microphysical processes associated to the monomer formation---especially the nucleation of initial condensates---are highly uncertain for exoplanetary atmospheres. Although the classical nucleation theory is available, as used in previous studies \citep[e.g.,][]{Helling&Fomins13,Powell+18,Gao&Benneke18}, one should keep in mind that the theory sometimes deviates from the nucleation rate measured by numerical and laboratory experiments by several orders of magnitudes \citep[e.g.,][]{Ford97,Tanaka+11,Lee+18}. 

In this study, we mimic the monomer formation by setting the size of monomers as a free parameter. For the sake of simplicity, every monomer is assumed to have the same size. We assume that the nucleation predominantly occurs at the cloud base, and the formed condensate particles instantaneously grow until all condensable vapor at the cloud base is incorporate into the particles. In other words, we calculate the growth of cloud particles in the region above which the monomer formation is completed (Figure \ref{fig:schematic}). %We discuss the validity and caveats of the assumption in Section \ref{sec:discussion}.}

%%%%%%%%%%%%%%%%%%%%%%%%%%
\subsubsection{Aggregate Growth and Transport above the Cloud Base}
% In this section, we briefly summarize the particle growth model \citep[for detail, see][and references therein]{Ohno&Okuzumi18}.
The formed monomers are collided each other and grow into the fluffy aggregates (Figure \ref{fig:schematic}). The aggregates are then mixed in the vertical direction by atmospheric circulation, which we approximate as a diffusion process in the horizontal averaged sense \citep{Parmentier+13,Charnay+15,Zhang&Showman18a,Zhang&Showman18b}.
The upward transport is limited by the downward settling motion of the particles. 
We treat these processes using 1D vertical transport equations with a collisional growth term \citep{Ohno&Okuzumi18}, 
\begin{equation}\label{eq:transport_nc}
\frac{\partial n_{\rm c}}{\partial t} = \frac{\partial}{\partial z}\left[ n_{\rm g}K_{\rm z}\frac{\partial }{\partial z}\left( \frac{n_{\rm c}}{n_{\rm g}}\right)+v_{\rm t}n_{\rm c} \right] - \left| \frac{\partial n_{\rm c}}{\partial t} \right|_{\rm coll},
\end{equation}
\begin{equation}\label{eq:transport_rhoc}
\frac{\partial \rho_{\rm c}}{\partial t} = \frac{\partial}{\partial z}\left[ \rho_{\rm g}K_{\rm z}\frac{\partial }{\partial z}\left( \frac{\rho_{\rm c}}{\rho_{\rm g}} \right)+v_{\rm t}\rho_{\rm c} \right],
\end{equation}
where $n_{\rm g}$ is the atmospheric gas number density and $|\partial n_{\rm c}/\partial t|_{\rm coll}$ is the decrease of the aggregate number density due to collisional growth.
We use the eddy diffusion coefficient $K_{\rm z}$ for GJ1214b derived by \citet{Charnay+15}:
\begin{equation}
K_{\rm z}=K_{\rm 0}\left(\frac{P}{1~{\rm bar}}\right)^{-2/5},
\end{equation}
where $K_{\rm 0}$ is the eddy diffusion coefficient at $1~{\rm bar}$ depending on the atmospheric metallicity, as listed in Table \ref{table:1}.

Collisional growth is induced by Brownian motion (coagulation hereafter) and differential gravitational settling (coalescence hereafter). We write $|{\partial n_{\rm c}}/{\partial t}|$ as 
\begin{equation}
\left| \frac{\partial n_{\rm c}}{\partial t} \right|_{\rm coll}=\left| \frac{\partial n_{\rm c}}{\partial t} \right|_{\rm coag}+\left| \frac{\partial n_{\rm c}}{\partial t} \right|_{\rm coal},
\end{equation}
where the first and second terms represent the contribution from coagulation and coalescence, respectively.
One can apply the same formula of collisional growth terms for spheres to aggregates by using the characteristic radius of aggregates \citep[e.g.,][]{Gao+17b}.
Approximating an aggregate with a sphere of characteristic radius $r_{\rm agg}$, the two terms can be written as 
\citep[e.g.,][]{Rossow78}
\begin{equation}\label{eq:dcoag}
    \left| \frac{\partial n_{\rm c}}{\partial t} \right|_{\rm coag}  =  {\rm min}\left( 8\sqrt{\frac{\pi k_{\rm B}T}{m_{\rm agg}}}r_{\rm agg}^2n_{\rm c}^2~,~ \frac{4k_{\rm B}T\beta}{3\eta}n_{\rm c}^2\right)
\end{equation}
and
\begin{equation}\label{eq:dcoal}
 \left| \frac{\partial n_{\rm c}}{\partial t} \right|_{\rm coal} \approx 2\pi r_{\rm agg}^2 n_{\rm c}^2 \epsilon v_{\rm t}E,
\end{equation}
where $\epsilon=0.5$ is the numerical factor already introduced in Section~\ref{sec:kataoka} and $E$ is the collection efficiency originated from the fact that an aggregate strongly coupled to the ambient gas cannot collide with another aggregate. We use the expression \citep{Guillot+14}
\begin{equation}
    E={\rm max}[0,1-0.42{\rm St}^{-3/4}],
\end{equation}
where ${\rm St}\equiv(v_{\rm t}/g)/(r_{\rm agg}/\epsilon v_{\rm t})$ is the Stokes number.
%Note that the collisional growth is mainly driven by the coagulation in our fiducial

%%%%%%%%%%%%%%%%%%%%%%%%%%%%
\subsubsection{Numerical Procedures}\label{sec:numerical}
We consider that the cloud particles are {in solid form and} made of pure KCl, which is a major condensable species formed in warm ($T=500$--$1000~{\rm K}$) exoplanets \citep{Morley+13,Lee+18}.
{For super-Earths, the pressure and temperature at cloud forming region \citep[$\la{0.1}~{\rm bar}$ and $\la{900}~{\rm K}$, see e.g.,][]{Gao&Benneke18} are well below the triple-point pressure and temperature of KCl \citep[${140}~{\rm bars}$ and ${1041}~{\rm K}$,][]{Rodrigues+07}.
Thus, the KCl clouds are likely made of solid particles that could grow into an aggregate.
}
% and studied by previous studies \citep{Morley+13,Morley+15,Charnay+15,Gao+18,Ohno&Okuzumi18,Gao&Benneke18}.
We suppose a hypothetical planet that has the PT profile and surface gravity ($g=8.93~{\rm m~s^{-2}}$) of GJ1214b.
The PT profile is calculated by an analytical model of \citet{Guillot10Adam} for cloud-free atmospheres as applied in \citet{Ohno&Okuzumi18}, but we additionally include the effect of the convective adjustment by setting the adiabatic lapse late $g/c_{\rm p}$ as an upper limit of a temperature gradient.

%We assume that each monomer is made of pure KCl.
%The surface energy of KCl is used for the calculation of the equillibrium filling factor and given by \citep{Janz&Dijkhuis69}
%\begin{equation}
%\gamma_{\rm KCl}=160.4-0.07T({\rm K})~{\rm erg~{cm}^{-2}}.
%\end{equation}
%This value is measured for a molten KCl.
%Strictly speaking, one should use the surface energy for a solid KCl; however, the results would not be seriously affected because the equilibrium filling factor is insensitive to the surface energy as shown in Section \ref{sec:porosity_recipe}.

To obtain the vertical profiles of $\rho_{\rm c}$ and $n_{\rm c}$, we solve Equations \eqref{eq:transport_rhoc} and \eqref{eq:transport_nc} until the system reaches a steady state.
The sizes of aggregates are calculated by using the equilibrium filling factor from Equation \eqref{eq:phi_eq} at each time step.
We note that the collisional compression should occur only when the particle collisions dominate over the vertical transport. 
Otherwise, the compression can occur without collision, which is clearly unrealistic.
To take into account it, we switch off the collisional compression if the vertical mixing timescale $\tau_{\rm mix}{\equiv}H^2/K_{\rm z}$ is shorter than the collisional growth timescale $|d\log{n_{\rm c}}/dt|^{-1}$.
%Collisional compression is generally negligible as shown in next section, but we discuss the impacts on the vertical structures in Section \ref{sec:discussion}.
The upper boundary condition is set to zero-flux, while the flux at the lower boundary is calculated assuming that $n_{\rm c}/n_{\rm g}$ and $\rho_{\rm c}/\rho_{\rm g}$ are constant at the cloud base. 
Since we have assumed that all condensable vapor is incorporated in the cloud particles at the cloud base (Section \ref{sec:nucleation}), the cloud mass density at the lower boundary is given by
\begin{equation}\label{eq:rhoc_0}
    %\rho_{\rm c}(P_{\rm b})=\rho_{\rm g}(P_{\rm b}),
    \rho_{\rm c}(P_{\rm b})=\rho_{\rm s}(P_{\rm b}),
\end{equation}
where $P_{\rm b}$ is the cloud-base height in pressure and $\rho_{\rm s}$ is the saturation vapor density of KCl, which is calculated by the saturation vapor pressure described in \citet{Morley+12}. For a given monomer radius, the number density of cloud particles at the lower boundary is also calculated as
\begin{equation}\label{eq:n_0}
n_{\rm c}(P_{\rm b})=\frac{3\rho_{\rm s}(P_{\rm b})}{4\pi r_{\rm mon}^3 \rho_{\rm p}}.
\end{equation}

%{\it Please merge this sentence with the sentences around Eqs. 32 and 33}
%The cloud base is the height in which the partial pressure of condensing vapor $q_{\rm v}P$, where $q_{\rm v}$ is the volume mixing ratio of the condensing vapor, exceeds the saturation vapor pressure $P_{\rm s}$. 
The top and bottom of the computation domain are imposed at ${10}^{-8}~{\rm bar}$ and the cloud-base height, which is determined by the volume mixing ratio of KCl vapor $q_{\rm v,KCl}$ listed in Table \ref{table:1} and the saturation vapor pressure.
The vertically coordinate $z$ is discretized into linearly spaced bins, depending on the atmospheric metallicity (Table \ref{table:1}).
The time increment is calculated at each time step so that the fractional decrease of $n_{\rm c}$ does not exceed $0.5$, i.e., $\Delta t \leq 0.5\times |\partial \log{n_{\rm c}}/\partial t|^{-1}$.

%We determine the number and mass densities of cloud particles at the cloud base as follows.
%When the most of vapor is condensed into cloud particles, usually achieved because of a very short condensation timescale, the mass mixing ratio of cloud particles is determined by the mixing ratio of a condensing vapor \citep{Ohno&Okuzumi18}.
%Therefore, the cloud mass density at the cloud base is given by
%\begin{equation}\label{eq:rhoc_0}
%    \rho_{\rm c}(P_{\rm b})=\rho_{\rm s}(P_{\rm b}),
%\end{equation}
%where $\rho_{\rm s}$ is the saturation vapor density of condensing vapor.
%For given monomer mass $m_{\rm mon}$, the number density of cloud particles at the cloud base is also given by  
%\begin{equation}\label{eq:n_0}
%n_{\rm c}(P_{\rm b})=\frac{\rho_{\rm s}(P_{\rm b})}{m_{\rm mon}}.
%\end{equation}
%Equation \eqref{eq:rhoc_0} and \eqref{eq:n_0} define the initial number and mass densities of cloud particles at the cloud base corresponding to the lower boundary of the computation domain.

\begin{table}[t]
  \centering
   \caption{Fiducial Parameters of This Study}
  \begin{tabular}{c|cccc}
     metallicity & $\mu_{\rm g}$ & $q_{\rm v, KCl}$ (mol/mol) & $K_{\rm 0}~({\rm m^{2}~s^{-1}})$ & $\Delta z$~(km) \\ \hline \hline
     1$\times$solar &  2.3 &1.83$\times {10}^{-7}$ & $7.0\times {10}^2$&20\\ 
    10$\times$solar & 2.5 & 1.80$\times {10}^{-6}$ & $2.8\times {10}^3$&20\\ 
        100$\times$solar & 4.3 & 1.70$\times {10}^{-5}$ & $3.0\times {10}^3 $&10\\
    1000$\times$solar  & 16.7  & 1.20$\times {10}^{-4}$ & $3.0\times {10}^2 $ & 5 
  \end{tabular}
  \label{table:1}
\end{table}
%%%%%%%%%%%%%%%%%%%%%%%%%%%%%%%%%%%%%

%%%%%%%%%%%%%%%%%%%%%%%%%%%%%%%%%%%%%%
\subsection{Results}\label{sec:result}
%%%%%%%%%%%%%%%%%%%%%%%%%%%%%%%%%%%%%%
%%%%%%%%%%%%%%%%%%
\begin{figure*}[t]
\includegraphics[clip,width=\hsize]{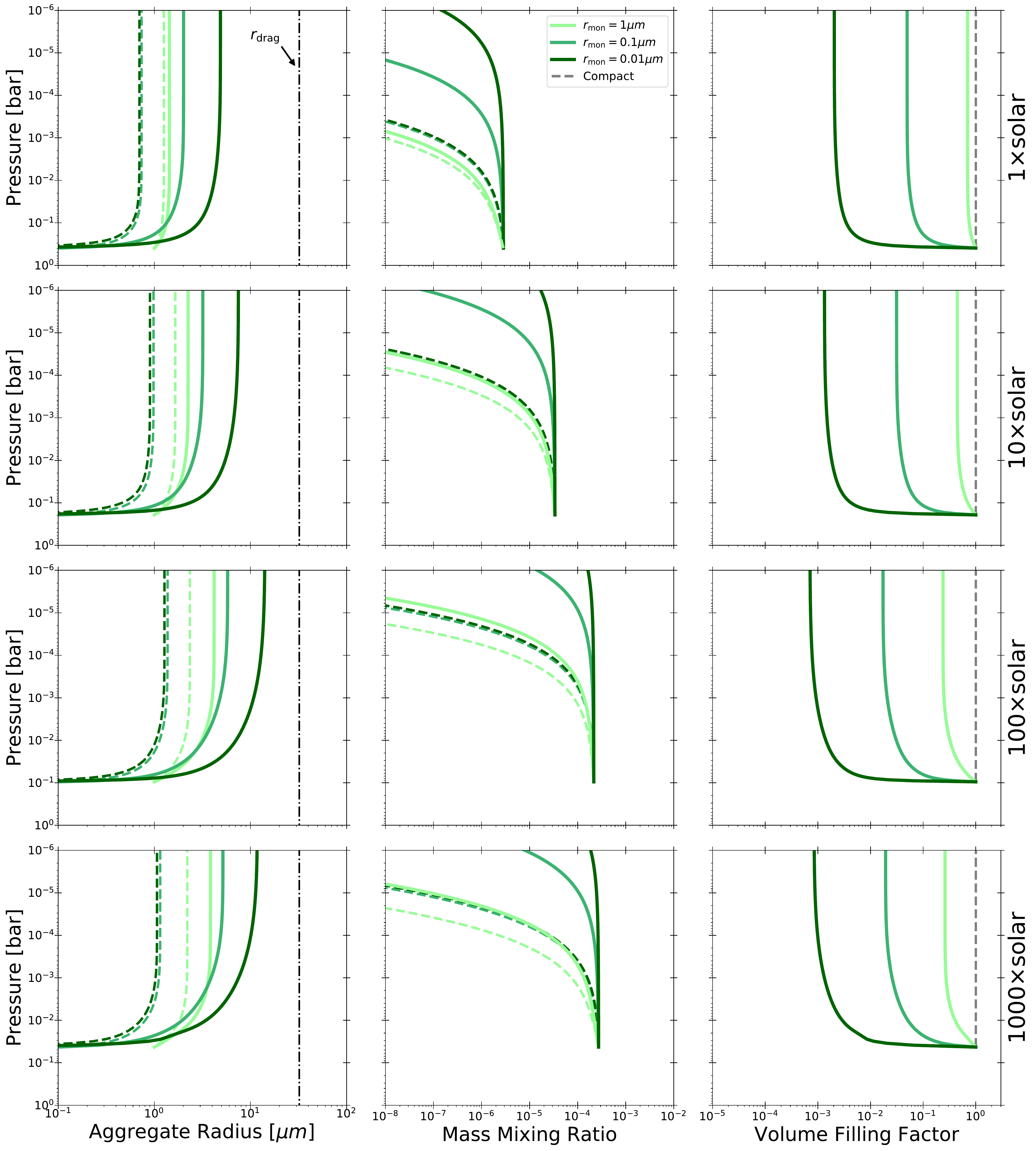}
\caption{
Vertical structure of a KCl cloud in GJ1214b from compact and fluffy aggregate models. The left, center, and right columns show the radius $r_{\rm agg}$, mass mixing ratio $ \rho_{\rm c}/\rho_{\rm g}$, volume filling factor of CPAs, respectively. The top, middle, and bottom rows are for atmospheric metallicities of $1\times$, $10\times$, $100\times$, and $1000\times$ solar, respectively. The vertical axes are atmospheric pressure for all panels. The light-green, green, and dark-green lines show the profiles for $r_{\rm mon}=1$, $0.1$, and $0.01~{\rm \mu m}$, respectively. The dotted lines also show the profiles for compact-sphere clouds ($D_{\rm f}=3$) for reference. The black dash-dot lines in the left column denote the compression radius $r_{\rm drag}$ given by Equation \eqref{eq:r_comp}.
}\label{fig:vertical}
\end{figure*}
%%%%%%%%%%%%%%%%%%
% \subsection{Influences of Porosity Evolution on Vertical Structure}\label{sec:vertical}
% In this section, we examine how the fluffy aggregates grows in exoplanetary atmospheres.
We here demonstrate how the porosity evolution affects the vertical profiles of KCl clouds.
Figure \ref{fig:vertical} shows the vertical distribution of the size $r_{\rm agg}$, cloud mass mixing ratio $q_{\rm c}=\rho_{\rm c}/\rho_{\rm g}$, and filling factor of aggregates $\phi_{\rm eq}$ for various monomer sizes and atmospheric metallicities.
We also plot the vertical profiles of compact ($D_{\rm f}=3$) sphere clouds for comparison.
The left panels of Figure \ref{fig:vertical} show that the cloud particles produced at the cloud base grow locally until the timescale of collisional growth becomes comparable to the vertical diffusion timescale. Well above the cloud base, no appreciable growth occurs because the growth timescale increases with height \citep{Ohno&Okuzumi18,Powell+18,Gao&Benneke18}.
Notably, the cloud mass mixing ratio for submicron monomer cases is high even at a very high altitude of $P < 10^{-4}~\rm bar$ as compared to the case of compact-sphere clouds.
%porous aggregates made of submicron-sized monomers can ascend to very high altitude of $P < 10^{-4}~\rm bar$, which are never reached by compact aggregates. 
The reason for this will be explained in a later part of this section.

We note that the aggregate sizes in upper atmospheres may decrease with height in reality, as seen in other studies \citep{Gao&Benneke18,Ormel&Min19}.
The trend is not captured in our calculations where the particle sizes are constant at upper atmospheres. 
This is caused by the fact that our model assumes a narrowly-peaked size distribution that cannot handle the decrease of the mean size caused by the removal of the largest particles from the distribution.
However, the size-decreasing effect is presumably not crucial for slowly settling CPAs.
This is because the effective size becomes nearly constant in vertical, as seen in our calculations, when the particles have sufficiently small sizes and thus small settling velocity \citep[see e.g., Figure 4 of][]{Gao+18}.

The trend of vertical size distribution is appreciably different between compact-sphere and fluffy-aggregate cases. For compact-sphere case, the particle size well above the cloud top decreases with decreasing monomer size $r_{\rm mon}$ because a higher number density at the cloud base (which corresponds to a smaller monomer size at the base; see Equation \ref{eq:n_0}) leads to a smaller particle size above the base \citep{Gao+18,Ohno&Okuzumi18,Ormel&Min19}.
{The trend is originated from the fact that, for a given mass mixing ratio, a total amount of condensing materials on each particle decreases with increasing a number density.
The coagulation is effective for a high number density, but halted once the particle size exceeds the threshold above which the number density becomes too low to cause the collisions \citep[see Section 3.2 of][]{Ohno&Okuzumi18}. }
By contrast, for fluffy-aggregate clouds, the aggregate radius at high altitude {\it increases} with decreasing monomer radius $r_{\rm mon}$.
As shown below, this is because the coagulation timescale is a function of aggregate mass and because aggregates made of smaller monomers have to grow to larger in size to obtain a certain mass.  
For aggregates larger than the mean free path of themselves, the timescale of coagulation growth $\tau_{\rm coag}\equiv|d\log{n_{\rm c}}/dt|_{\rm coag}^{-1}$ is approximately given by
\begin{equation}
\tau_{\rm coag}\approx \frac{3\eta}{4k_{\rm B}Tn_{\rm c}},
\end{equation}
which follows from Equation~\eqref{eq:dcoag}.
% where we assume the diffusive coagulation.
Using the relation $q_{\rm c}\rho_{\rm g}=m_{\rm agg}n_{\rm c}$, we obtain
\begin{equation}
\tau_{\rm coag}\approx \frac{3\eta m_{\rm agg}}{4k_{\rm B}T\rho_{\rm g}q_{\rm c}},
\end{equation}
which indicates that the coagulation timescale is independent of aggregates properties other than $m_{\rm agg}$.
Since the final size is determined by the balance between coagulation and mixing timescales ($\tau_{\rm coag}= \tau_{\rm mix}$), the final aggregate mass is given by
\begin{equation}\label{eq:r_agg_fin}
m_{\rm agg} \approx \frac{4k_{\rm B}TH^2}{3\eta K_{\rm z}}\rho_{\rm g}(P_{\rm *})q_{\rm c},
\end{equation}
where $P_{\rm *}$ is the pressure level where the growth is completed.
For $D_{\rm f} = 2$, the aggregate mass scales as $m_{\rm agg}\propto r_{\rm agg}^2r_{\rm mon}$, and hence the final aggregates radius increase with decreasing monomer size.

The aggregate size slightly increase with increasing on atmospheric metallicity.
In the case of $r_{\rm mon}=0.1~{\rm \mu m}$, for example, the aggregate radii at high altitude are $2$, $3$, $5$, and $5~{\rm \mu m}$ for the metallicities of $1\times$, $10\times$, $100\times$, and $1000\times$ solar abundance, respectively.
The increase of the aggregate size is caused by the fact that a higher atmospheric metallicity ($q_c$ at the cloud base) leads to a higher cloud density that facilitates coagulation growth.
This can also be seen from Equation \eqref{eq:r_agg_fin}, which shows $m_{\rm agg}\propto q_{\rm c}$. 
However, the aggregate size also depends on the mixing timescale  $H^2/K_{\rm z}$ (see Equation \ref{eq:r_agg_fin}), which decreases with increasing the atmospheric metallicity in our parameter set. This effect substantially cancels out the effects of $q_{\rm c}$, which explains the weak metallicity-dependence of the aggregate size in Figure \ref{fig:vertical}.
%The aggregate size is also influenced by the metallicity dependence of $H^2/K_{\rm z}$, but the effect is relatively minor compared to the dependencies of $q_{\rm c}$.
%We note however that the metallicity dependence of $K_{\rm z}$ and $H$ also affects the final aggregate mass to some extent \edit1{\it Are their effects minor?}.

The key result of this section is that the aggregates never experience compression in the cases studied here. 
The dot-dashed lines in the left panels of Figure~\ref{fig:vertical} show the threshold size for the gas-drag compression $r_{\rm drag}$ (Equation \ref{eq:r_comp}) above which the aggregates leave fractal growth. 
Figure \ref{fig:vertical} shows that the particle growth is insufficient to reach the threshold size for the gas-drag compaction. Although the collisional compression can operate on micron-size aggregates in upper atmospheres ($P\la{10}^{-3}~{\rm bar}$, see Equation \ref{eq:r_coll}), it does not take place there because the number density is too low to cause the particle collision, i.e., $\tau_{\rm coll}\gg\tau_{\rm mix}$. As a result, aggregates are fractal ($D_{\rm f} \approx 2$) even at high altitude.
%The collisional compression never occurs because the larger monomer size, which is needed to cause the compression (Section \ref{sec:kataoka}), leads to lower number density (Equation \ref{eq:n_0}) and thus less efficient particle growth.}
%The gas-drag compression only occurs in the case of $r_{\rm mon}=0.01~{\rm \mu m}$ for metallicities higher than $100\times$ solar. 
%As a result, aggregates are fractal ($D_{\rm f} \approx 2$) even at high altitude. 

The absence of the compression enables us to evaluate the vertical extent of clouds. The cloud particle aggregates can ascend to the height of $\tau_{\rm mix}{\sim}\tau_{\rm fall}$, where $\tau_{\rm fall}\equiv H/v_{\rm t}$ is the falling timescales \citep[e.g.,][]{Charnay+15}. Assuming $l_{\rm g}{\gg}r_{\rm agg}$ for upper atmospheres, the terminal velocity can be approximated as
\begin{equation}\label{eq:vt_BCCA}
v_{\rm t} \approx \frac{\rho_{\rm mon}g}{\rho_{\rm g}v_{\rm th}}r_{\rm mon},
\end{equation}
where we use the relation $r_{\rm agg} \rho_{\rm agg} = r_{\rm mon}\rho_{\rm mon}$ for $D_{\rm f}=2$. Solving $\tau_{\rm mix}=\tau_{\rm fall}$ about the pressure, we find the pressure level $P_{\rm top}$ to which cloud particles can ascend: 
\begin{eqnarray}\label{eq:Ptop}
P_{\rm top}&\approx&\frac{\rho_{\rm mon}g^2H^2r_{\rm mon}}{v_{\rm th}K_{\rm z}}\\
\nonumber
&\sim&0.03~{\rm mbar} \left( \frac{r_{\rm mon}}{0.1~{\rm \mu m}}\right)  \left( \frac{\rho_{\rm mon}}{2~{\rm g~{cm}^{-3}}}\right) \left( \frac{K_{\rm z}}{10^{4}~{\rm m^2~{s}^{-1}}}\right)^{-1} \left( \frac{v_{\rm th}}{1~{\rm km~{s}^{-1}}}\right)^{3},
\end{eqnarray}
where we use $v_{\rm th}=\sqrt{(8/\pi)gH}$. Equation \eqref{eq:Ptop} indicates that $P_{\rm top}$ is independent of the size of cloud particle aggregates $r_{\rm agg}$. This explains why the cloud particle aggregates made of smaller monomers can ascend higher altitude in Figure \ref{fig:vertical} despite their very large sizes ($\gg 1~{\rm \mu m}$).

\section{Transmission Spectrum with Fluffy-Aggregate Clouds}\label{sec:result2}
%%%%%%%%%%%%%%%%%%%%%%%%%%%%%%

The optical properties of fluffy aggregates are considerably different from those of compact spheres.
In addition, fluffy aggregates are able to ascend to very high altitude as demonstrated in Section \ref{sec:vertical}.
In this section, we investigate how these effects influence the transmission spectra of exoplanets.
% its optical properties would significantly affect the results of observations, especially transmission spectra.
% The distinct optical properties might produce new interpretations of observed spectra and offer the observable signature of fluffy aggregates.
% To investigate these topics, we calculate synthetic transmission spectra with the fluffy-aggregate clouds.

%%%%%%%%%%%%%%%%%%%%
\subsection{Method}
%%%%%%%%%%%%%%%%%%%%

We calculate synthetic transmission spectra of GJ1214b, a super-Earth believed to be covered by clouds (and/or hazes) in very high altitude \citep[e.g.,][]{Kreidberg+14}, using the vertical profiles of KCl clouds obtained in Section \ref{sec:vertical}.
We do this by calculating the wavelength-dependent transit depth $D(\lambda)$ of a planet, which can be expressed as \citep[e.g.,][]{Heng&Kitzmann17}
\begin{equation}\label{eq:D}
D(\lambda)=\frac{\pi R_{\rm 0}^2 + 2\pi \int_{R_{\rm 0}}^{\infty} [1-\exp{(-\tau_{\rm s})}]rdr}{\pi R_{\rm *}^2},
\end{equation}
where $R_{\rm 0}$ is the reference transit radius and $\tau_{\rm s}$ is the optical depth for slant viewing geometry, called the slant optical depth, and $r$ is the distance from the center of the planet.
We take $R_0$ to be the radius at the pressure level of $10~{\rm bar}$ following previous studies \citep[e.g.,][]{Kreidberg+15}. The slant optical depth $\tau_{\rm s}$ is calculated by integrating the extinction by gas molecules and cloud particles along the observer's line of sight \citep[e.g.,][]{Fortney+03}:
\begin{equation}
\tau_{\rm s}(r)=2\int_{ r}^{\rm \infty} (\alpha_{\rm g}+\alpha_{\rm c})\frac{r'dr'}{\sqrt{r'^{2}-r^2}},
\end{equation}
where $\alpha_{\rm g}$ and $\alpha_{\rm c}$ are the extinction efficiencies of gas molecules and cloud particles, respectively.
The stellar radius $R_*$ and planet's semi-major axis $a$ are taken to be $R_{\rm *}=0.216R_{\rm sun}$ and $a=0.014~{\rm au}$, which are the values for GJ1214b from the Exoplanet eu catalog \footnote{http://exoplanet.eu}.

%%%%%%%%%%%%%%%%%%%%%%%%%
\subsubsection{Gas Opacity}
%%%%%%%%%%%%%%%%%%%%%%%%%
To evaluate the gas opacity, we calculate the mixing ratio of gas molecules using the open-source Thermochemical Equilibrium Abundances (TEA) code \citep{Blecic+16}. 
The TEA calculates the gas mixing ratio in thermochemical equilibrium for given temperature, pressure, and elemental abundances based on \citet{Asplund+09} using the Gibbs free-energy minimization method.
Following \citet{Freedman+08,Freedman+14}, we take into account the molecular absorption of  $\rm H_{\rm 2}$, $\rm H_{2}O$, $\rm CH_{4}$, $\rm CO$, $\rm CO_{2}$, $\rm NH_{3}$, $\rm H_{2}S$, and $\rm PH_{3}$ as well as the Rayleigh scattering of the molecules.
We calculate the absorption and scattering cross sections of the molecules following the method of \citet{Kawashima&Ikoma18} with the line list of HITRAN2016.
The Voigt function is calculated by the polynomial expansion \citep{Kunz97,Ruyten04}, and the total internal partition function sums are calculated by TIPS code \citep{Gamache+17}.
We refer readers to the relevant literature
\citep[e.g.,][]{Rothman+98,Sharp&Burrows07,Malik+19} for detail methodology of the gas opacity calculations.
Further improvements of the line lists and the broadening coefficients \citep[e.g.,][]{Tennyson+18,Gharib-Nezhad&Line18} remain for future studies, as our current focus is to study how the fluffy-aggregate clouds influence the transmission spectra.

%%%%%%%%%%%%%%%%%%%%%%%%%%%%%%%%
\begin{figure*}
\includegraphics[clip,width=\hsize]{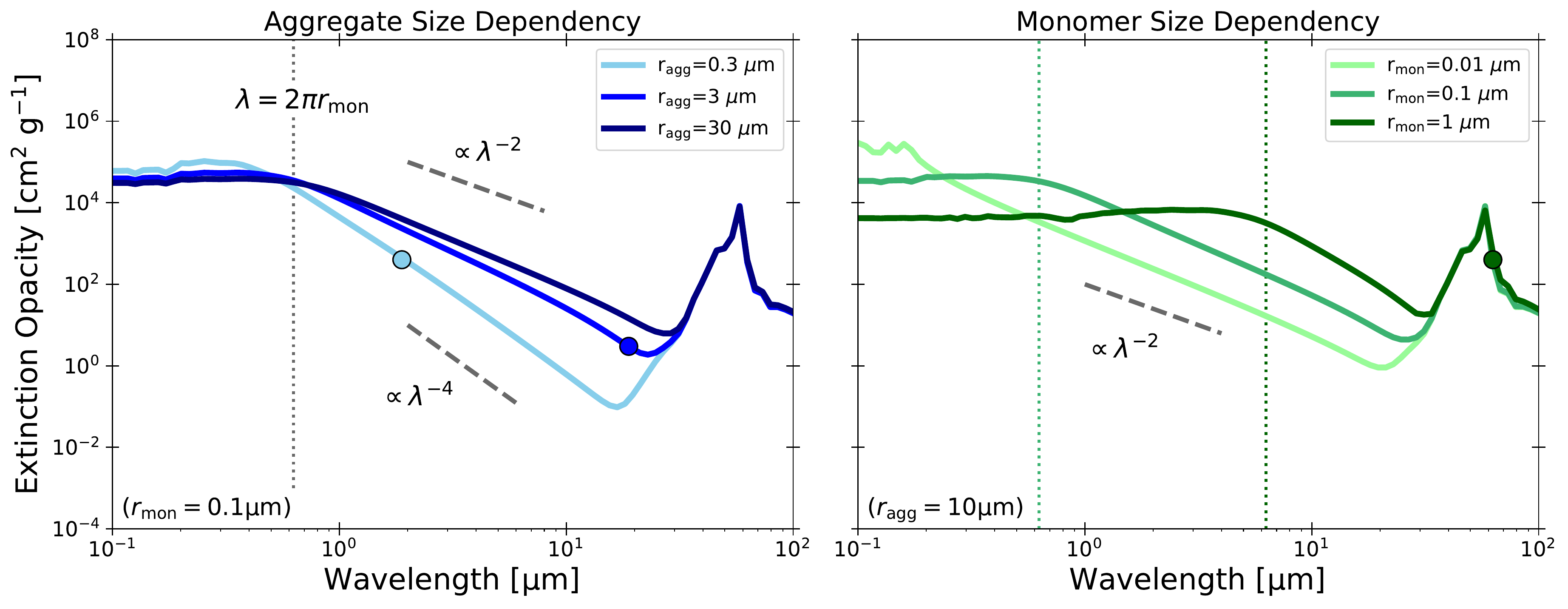}
\caption{Extinction opacity of KCl aggregates with $D_{\rm f}=2$ as a function of wavelength for different aggregate sizes $r_{\rm agg}$ and monomer sizes $r_{\rm mon}$, calculated by the MMF theory. The left panel is for aggregates of fixed $r_{\rm mon} = 0.1~\micron$ and different $r_{\rm agg}$, wheres the right panel is for fixed $r_{\rm agg} = 10~\micron$ and different $r_{\rm mon}$. The wavelength corresponding to $2\pi r_{\rm mon}$ and $2\pi r_{\rm agg}$ are denoted as dotted lines and filled circles, respectively.
 }\label{fig:opacity}
\end{figure*}
%%%%%%%%%%%%%%%%%%%%%%%%%%%%%%%%%%%
%%%%%%%%%%%%%%%%%%%%%%%%%%%%%%%
\subsubsection{Aggregates Opacity}\label{sec:kappa_agg}
%%%%%%%%%%%%%%%%%%%%%%%%%%%%%%%

The Mie theory \citep[e.g.,][]{Bohren&Huffman83} is usually used for the calculations of the  opacity of spherical particles, but is no longer valid for irregular aggregates.
The Mie theory coupled with the effective medium theory is one of the ways to calculate the aggregate opacity \citep[][]{Marley+13}.
However, this approach also fails to reproduce scattering properties of an aggregate when the relevant wavelength is much smaller than the aggregate \citep{Tazaki+16,Tazaki&Tanaka18}.
Aggregates potentially grow to $1$--$10~{\rm \mu m}$ in size as shown in Section \ref{sec:vertical}, while current and future observations mainly use shorter wavelengths such as $1.1$--$1.7~{\rm \mu m}$ for HST/WFC3, $0.6$--$5~{\rm \mu m}$ for JWST/NIRSpec \citep[][]{Batalha+17}, and $1.25$--$7.8~{\rm \mu m}$ for ARIEL \citep{Tinetti+16}.
Therefore, the effective medium theory is still not a good approximation especially for upcoming observations.%in the context of this study.

To properly calculate the aggregate opacity, we apply the modified mean field (MMF) theory \citep{Tazaki&Tanaka18}.
The MMF theory is based on the Rayleigh-Gans-Debye (RGD) theory that calculates the interference of single-scattered waves from every monomer by taking the aggregate structure into account \citep{Tazaki+16} with modifications for multiple scattering within an aggregate using the mean field assumption \citep[][]{Berry&Percival96}. 
%\edit1{\it The description of MMF is a little bit lengthy. Maybe the following sentences can be omitted: 
%If one adopts the mean field assumption in which every monomer receives same multiple scattering field \citep{Berry&Percival86,Botet+97}, the RGD theory can successfully treat the multiple scattering within the aggregate.
%After slight modifications, the RGD theory with the mean field assumption becomes the MMF theory \citep[for detail, see][]{Tazaki&Tanaka18}.}
The MMF theory successfully reproduces the extinction, absorption, and scattering opacities of aggregates calculated by the rigorous T-matrix method in a wide range of wavelength \citep[][]{Tazaki&Tanaka18}.
For calculations, we apply the Gaussian cut-off for the two-points correlation function that specifies the aggregate structure \citep{Tazaki+16}.

%In addition to the opacity calculation, we also take into account the reduction of effective opacity due to forward scattering \citep[e.g.,][]{Robinson17} using an analytical model of \citet{Robinson+17}, although the effect is almost negligible for parameters of GJ1214b.

%\textcolor{blue}{We note that the imaginary part of refractive index of KCl is nearly zero in $0.2$--$10~{\rm \mu m}$, and thus the absorption is only responsible to $\lambda<0.2~{\rm \mu m}$ and $>10~{\rm \mu m}$.}

Figure \ref{fig:opacity} shows the extinction opacity of KCl aggregates of $D_{\rm f}=2$ for different aggregate sizes $r_{\rm agg}$ and monomer sizes $r_{\rm mon}$. 
The refractive index of KCl is taken from \citet{Palik85} compiled by \citet{Kitzmann&Heng18}.
In the examples presented here, the extinction opacity is dominated by scattering in the wavelength range $\lambda \sim 0.2$--$10~\micron$.
At longer wavelengths, absorption dominates over scattering, and the absorption peak of KCl appears at $\lambda \sim 50~\micron$.
It is worth noting that the absorption feature is visible even if aggregate size is very large, as seen in the case of $r_{\rm agg}=30~{\rm \mu m}$.
This is because, unless the multiple scattering becomes dominant, the absorption cross section of an aggregate is the sum of the absorption of every monomer, and thus the wavelength dependence is the same as that of an individual monomer \citep{Berry&Percival96,Tazaki&Tanaka18}.

According to the MMF theory, the optical properties of an aggregate behave differently among three wavelength regimes $\lambda \ll 2\pi r_{\rm mon}$, $2\pi r_{\rm mon} \ll \lambda \ll 2\pi r_{\rm agg}$, and $\lambda \gg 2\pi r_{\rm mon}$. 
In the first regime, geometric optics applies to the constituent monomers, and the scattering cross section is approximately given by $\sigma_{\rm s} \sim \pi r_{\rm agg}^2$, independent of wavelength.
In the opposite limit of $\lambda \gg 2\pi r_{\rm agg}$, the Rayleigh limit applies to the aggregate, and the scattering cross section obeys the well-known law  $\sigma_{\rm s} \propto \lambda^{-4}$. 
I the left panel of Figure~\ref{fig:opacity},
this can be seen in the case of $r_{\rm agg} = 0.3~\micron$, at $\lambda \sim 1$--$10~\micron$.

The intermediate regime $2\pi r_{\rm mon}\ll \lambda \ll 2\pi r_{\rm agg}$
provides unique opacity properties for aggregates.
%aggregate that is not present for spherical particles.
For this regime, 
we find that the scattering opacity scales with wavelength dependence as $\sigma_{\rm s} \propto \lambda^{-2}$ (see Figure 4). In this intermediate regime, the scattered wave by an aggregate is a superposition of singly scattered waves from individual monomers, and the scattering cross section of a $D_{\rm f}=2$ aggregate has following dependence. \citep[][Section 5]{Berry&Percival96}
\begin{equation}\label{eq:sigma_s}
% \sigma_{\rm s}\propto \lambda^{-2}[\log{(4\pi^2 r_{\rm agg}^2k_{\rm 0})}-\log{(b\lambda^2)}],
\sigma_{\rm s}\propto r_{\rm agg}^2r_{\rm mon}^2\lambda^{-2}\log{(16\pi^2r_{\rm agg}^2/b\lambda^2)},
\end{equation}
where $b$ is a constant order of unity.
 %\edit1{\it Does the unwritten prefactor depends on monomer/aggregate sizes? Is there any strong reason that the prefactor is omitted?}.
%\textcolor{blue}{\it I added the unwritten dependencies of monomer/aggregate size.}
% where $b=[(D_{\rm f}+1)D_{\rm f}]^{1/2}$ and $\sigma_{\rm s,mon}$ is the scattering cross section of each monomer \edit1{\it where is $\sigma_{\rm s,mon}$?}
%the second term in the bracket of Equation \eqref{eq:sigma_s} \edit1{is} negligible, giving $\sigma_{\rm s} \propto \lambda^{-2}$.
This explains the scattering slope for $r_{\rm agg}=3~{\rm \mu m}$ and $30~{\rm \mu m}$ in the left panel of Figure \ref{fig:opacity}.
The unique scattering slope is caused by interference among the scattered waves from individual monomers.
The scattered waves toward large scattering angles ($\ga \lambda/2\pi r_{\rm agg}$) cancel out because of the presence of waves with opposite phases, leading to the $\lambda^{-2}$ dependence \citep{Kataoka+14}.
% \edit1{For $r_{\rm mon} \ga 0.1~\micron$, the opacity is nearly independent of wavelength at} $\lambda\la0.6~{\rm \mu m}$ and decreases with increasing wavelength at $\lambda\ga0.6~{\rm \mu m}$.
% The opacity at \edit1{ $\lambda \sim 1$--$10~{\rm \mu m}$ depends on aggregate size, with} $\kappa \propto \lambda^{-4}$ for $r_{\rm agg}=0.3~{\rm \mu m}$ and $\kappa \propto \lambda^{-2}$ for $r_{\rm agg}=3~{\rm \mu m}$ and $30~{\rm \mu m}$.
% The middle panel of Figure \ref{fig:opacity} shows the dependency of the monomer size, and one can confirm that the opacity for various monomer size are same independent of wavelength for $\lambda \ll 2\pi r_{\rm mon}$.
% For $\lambda \gg 0.6~{\rm \mu m}$, the dependency of $\propto \lambda^{-4}$ for $r_{\rm agg}=0.3~{\rm \mu m}$ is also equivalent to the Rayleigh scattering because the aggregate size parameter is smaller than unity (i.e., $\lambda\gg2\pi r_{\rm agg}$), which is also suggested by the effective medium theory \citep{Kataoka+14}.

%%%%%%%%%%%%%%%%%%%%%%%%%%%%%%%%%%%%
\subsection{Cloud-top Pressure}\label{sec:cloud-top}
%%%%%%%%%%%%%%%%%%%%%%%%%%%%%%%%%%%%%
\begin{figure*}[t]
\includegraphics[clip,width=\hsize]{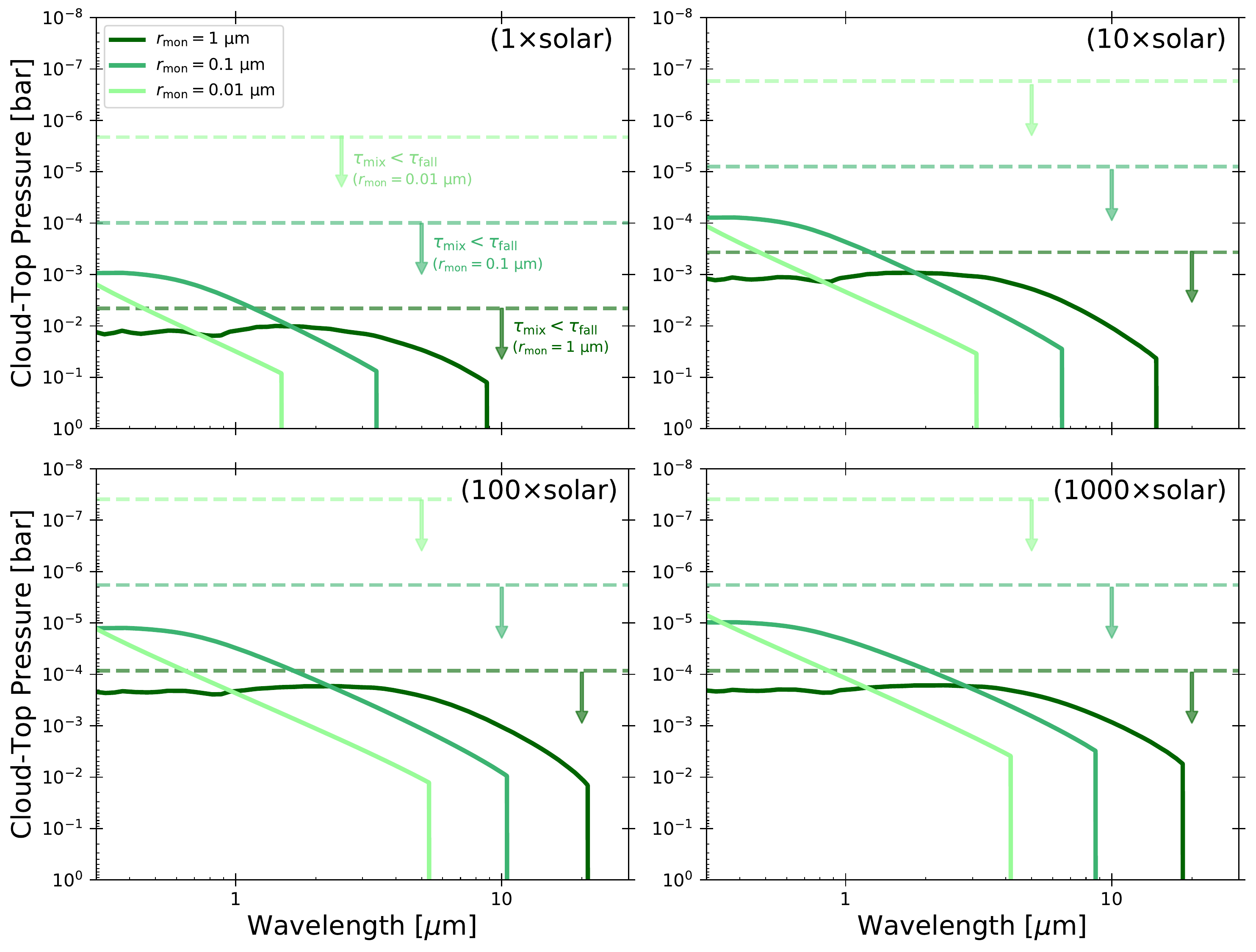}
\caption{
Cloud-top pressure of the fluffy-aggregate clouds as a function of wavelength. The dark-green, green, and light-green lines are for $r_{\rm mon}=1$, $0.1$, and $0.01~{\rm {\mu}m}$, respectively. The dashed lines indicate the pressure level of $\tau_{\rm mix}=\tau_{\rm fall}$ for each monomer size. Each panel exhibits the result for different atmospheric metallicity.}\label{fig:cloud_top}
\end{figure*}
%%%%%%%%%%%%%%%%%%%%%%%%%%%%%%%%%%%%%
{
Before showing the synthetic spectra, we investigate the cloud-top pressure, defined as the pressure level at which the cloud becomes optically thick along the line of sight of an observer (i.e., $\tau_{\rm s}=1$).
The cloud-top pressure clarifies the observable region of atmospheres and was examined by previous studies \citep[][]{Ohno&Okuzumi18,Powell+18,Gao&Benneke18,Helling+19}. 
Figure \ref{fig:cloud_top} shows the cloud-top pressure of fluffy-aggregate clouds as a function of wavelength for different monomer sizes and the atmospheric metallicities.
In general, the cloud top is located at a lower atmosphere for longer wavelength because the scattering opacity decreases with increasing wavelength for $\lambda>2\pi r_{\rm mon}$ (see Figure \ref{fig:opacity}).
We note that the cloud top hardly exceeds the altitude of $\tau_{\rm mix}=\tau_{\rm drag}$ (dashed lines in Figure \ref{fig:cloud_top}) for parameter ranges examined in this study.
{In near-infrared wavelength, the cloud-top height increases with decreasing monomer size as long as $r_{\rm mon}\ga0.1~{\rm \mu m}$, as CPAs constituted by smaller monomers ascend to higher altitude.
On the other hand, the cloud-top height also decreases with decreasing monomer size for $r_{\rm mon}\la0.1~{\rm \mu m}$.
This opposite trend is caused by the monomer size dependence of aggregate scattering opacity.
Using Equation \eqref{eq:sigma_s} and an aggregate mass $m_{\rm agg}\propto r_{\rm agg}^2r_{\rm mon}$, one can see that the scattering mass opacity follows
\begin{equation}
    \kappa_{\rm s}\equiv \frac{\sigma_{\rm s}}{m_{\rm agg}}\propto r_{\rm mon}\lambda^{-2}\log{(16\pi^2r_{\rm agg}^2/b\lambda^2)}.
\end{equation}
Thus, the scattering mass opacity decreases with decreasing monomer size.
On the other hand, in the limit of small monomer size (i.e., $\tau_{\rm mix}\ll\tau_{\rm fall}$), the cloud mass mixing ratio $q_{\rm c}$ is vertically constant and independent of monomer size (see Figure \ref{fig:vertical}).
Therefore, the scattering efficiency ($\alpha_{\rm c}=\rho_{\rm g}q_{\rm c}\kappa_{\rm s}$) and thus the cloud-top height decrease with decreasing monomer size for very small $r_{\rm mon}$.

}
%We find that smaller monomers lead to the cloud top at high altitude, as CPAs constituted by smaller monomers ascend to higher altitude.
%In the longer wavelength, the cloud-top pressure for large monomers are comparable to those for small monomers because scattering opacity increases with increasing a monomer size (see right panel of Figure \ref{fig:opacity}).
%Overall, the cloud-top pressure of the fluffy-aggregate clouds are higher than those of the compact-sphere clouds (dashed lines in Figure \ref{fig:cloud_top}).

We also find that the cloud-top pressure tends to be smaller for higher atmospheric metallicities.
This is because the cloud mass mixing ratio increases with increasing the metallicity.
Specifically, the cloud-top pressure for the atmospheric metallicity of $100$ and $1000\times$ solar reach $P\sim{10}^{-5}~{\rm bar}$ at near-infrared wavelength if the monomer is smaller than ${1}~{\rm \mu m}$.
It is worth pointing that the fluffy-aggregate clouds can produce the cloud top at the pressure level comparable to that retrieved from the observations of HST/WFC3 for GJ1214b \citep{Kreidberg+14}, which was hardly attained by the compact-sphere clouds in our previous study \citep{Ohno&Okuzumi18}.
}

%%%%%%%%%%%%%%%%%%%%%%%%%%%%%%%%%%%%
\subsection{Synthetic Spectra}\label{sec:spectrum}

%%%%%%%%%%%%%%%%%%%%%%%%%%%%%%%%%%%%%
\begin{figure*}[t]
\includegraphics[clip,width=\hsize]{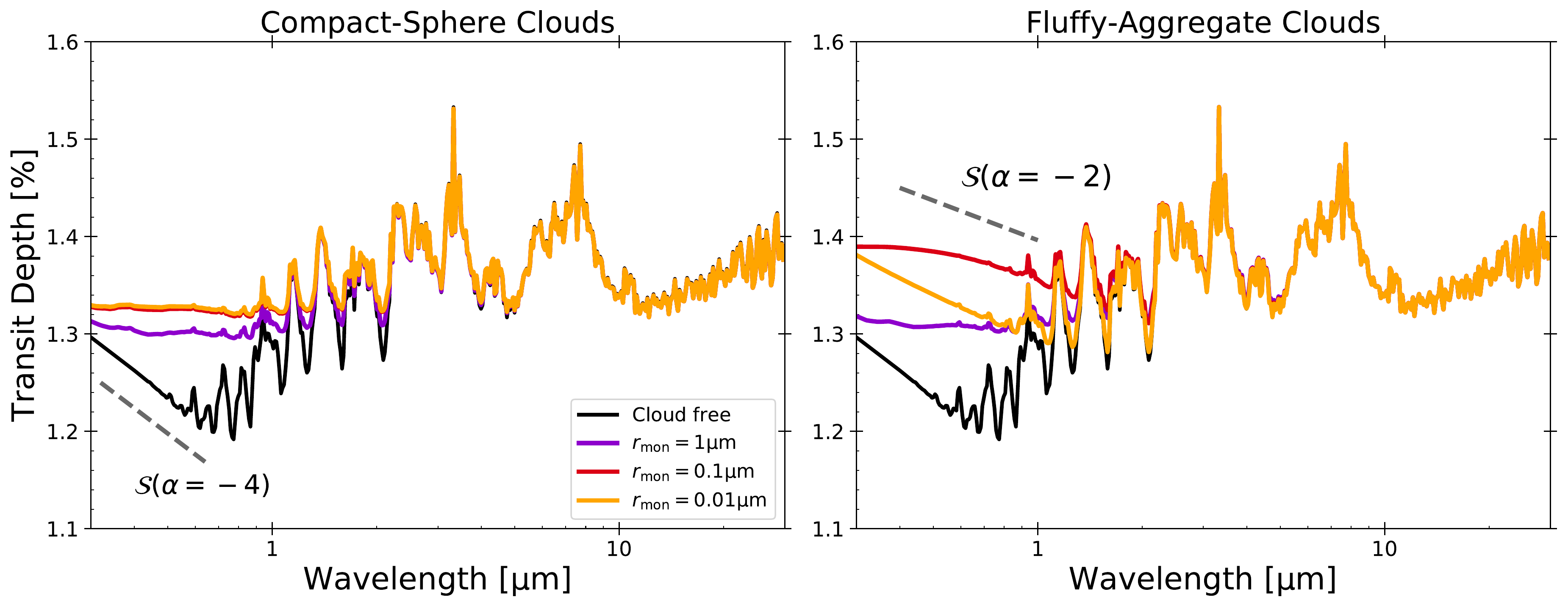}
\caption{
Synthetic transmission spectra of GJ1214b with a solar-metalicity atmosphere, from compact-{sphere} and fluffy-aggregate models (left and right panels, respectively) presented in Section~\ref{sec:vertical}. The purple, red, and orange lines are from the models assuming the monomer radii of $r_{\rm mon}=1~{\rm \mu m}$, $0.1~{\rm \mu m}$, and $0.01~{\rm \mu m}$, respectively.
In the compact-sphere models, the monomer size merely determines the number density of cloud particles at the cloud base (see Equation \ref{eq:n_0}).
For comparison, the spectrum for a cloud-free atmosphere is also shown by the black line.
The gray dashed lines denote the spectral slopes corresponding to $\alpha{\propto}\lambda^{-4}$ for the left panel, and ${\propto}\lambda^{-2}$ for the right panel (see Equation \ref{eq:slope}).
For clarity, the spectral resolution is binned down to $\lambda/\Delta \lambda \approx 100$, corresponding to the resolution of HST/WFC3.}\label{fig:spectrum_com}
\end{figure*}
%%%%%%%%%%%%%%%%%%%%%%%%%%%%%%%%%%%%%
%%%%%%%%%%%%%%%%%%%%%%%%%%%%%%%%%%%%%
\begin{figure*}[t]
\includegraphics[clip,width=\hsize]{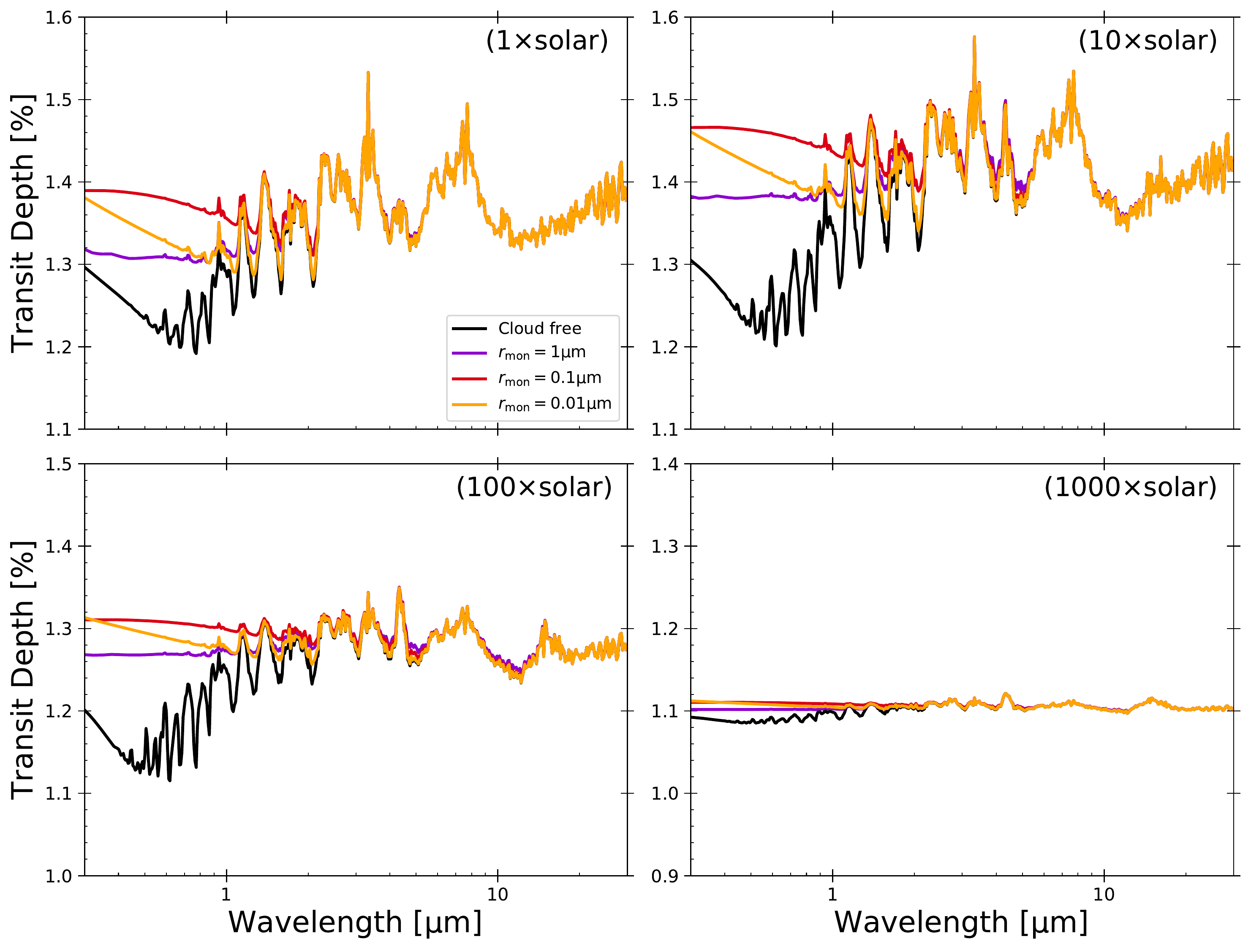}
\caption{Synthetic transmission spectra of GJ1214b with a cloud of fluffy KCl aggregates for various atmospheric metallicities. 
% Each panel is same as Figure \ref{fig:spectrum_com} but the atmospheric metallicity is varied. 
%\edit1{{\it This shoudn't be here:} We set  $R_{\rm 0}=2.25R_{\rm Earth}$ for every panels.} }\label{fig:spectrum_met}
}\label{fig:spectrum_met}
\end{figure*}
%%%%%%%%%%%%%%%%%%%%%%%%%%%%%%%%%%%%%

We begin by studying how the aggregate structure affects transmission spectra.
For later convenience, we introduce a metric characterizing the spectral slope, given by \citep[e.g.,][]{Line&Parmentier16}
\begin{equation}\label{eq:slope}
\mathcal{S}\equiv\frac{dD(\lambda)}{d\log{\lambda}}=\frac{2\pi R_{\rm p}H}{\pi R_{\rm *}^2}\alpha,
\end{equation}
where $\alpha$ is the power-law index of the extinction efficiency of atmosphere, i.e., ($\alpha_{\rm g}+\alpha_{\rm c})\propto \lambda^{\alpha}$. For example, $\alpha=-4$ for the Rayleigh scattering particles, and $\alpha=0$ for gray cloud particles.
Here, we have naively used the pressure scale height $H$ instead of the cloud scale height. 
{Strictly speaking, the {the could scale height is equal to $H$} only when the particle settling timescale is much longer than the mixing timescale \citep[see e.g., Equation (33) of][]{Ohno&Okuzumi18}. 
{The cluod scale height is smaller than $H$}  at high altitude where the cloud mass mixing ratio decreases with increasing height, implying $\tau_{\rm fall}\la \tau_{\rm mix}$.
{However, cloud particles at such very high altitude are usually so depleted that their contribution to transmission spectra is small. 
{In fact, as shown in previous section, the cloud top hardly exceeds the the pressure level of $\tau_{\rm mix}=\tau_{\rm fall}$ for the parameter space examined in this study.}
Therefore, Equation \eqref{eq:slope} offers a reasonable diagnosis of the spectral slope.
}
% Nevertheless, Equation \eqref{eq:slope} offers a reasonable diagnosis of the spectral slope.
% This is because the mass mixing ratio is very low at the region where the deviation occurs, and thus the extinction caused by the cloud is too weak to influence the shape of the spectrum.
}
%because both of them are identical if the particle settling timescale is much longer than the mixing timescale \citep[see e.g., Equation (33) of][]{Ohno&Okuzumi18}, expected for fluffy-aggregate clouds.

Figure \ref{fig:spectrum_com} shows the synthetic transmission spectra of GJ1214b with a solar-metallicity atmosphere and with a KCl cloud obtained from compact-{sphere} and fluffy-aggregate models (see the top rows of Figure~\ref{fig:vertical} for the cloud vertical structure).
We set the reference radius to $R_{\rm 0}=2.25R_{\rm Earth}$ so that the cloud-free solar-composition atmosphere produces the planet-to-star radius ratio of $R_{\rm p}/R_{\rm *}{\sim}0.115$ (i.e., $D{\sim}1.3\%$) in near-infrared \citep[e.g.,][]{Narita+13a}.
We calculate the optical properties of compact {spheres} using the Mie theory \citep[e.g.,][]{Bohren&Huffman83}.
For comparison, we also plot the transmission spectrum for the cloud-free atmosphere, which exhibits molecular absorption signatures of mainly $\rm H_2O$ molecules and the spectral slope in $\lambda \la0.5~{\rm \mu m}$ caused by the Rayleigh scattering of $\rm H_2$ molecules.
%The spectrum for the cloud-free atmosphere exhibits a lot of molecular absorption features, mainly produced by $\rm H_{\rm 2}O$ molecules, as well as the so-called Rayleigh slope in the visible wavelength ($\lambda \la0.5~{\rm \mu m}$) caused by hydrogen molecules.
In the left panel of Figure \ref{fig:spectrum_com}, the compact-{sphere clouds} produce a floor of the transit depth at $\lambda \la 2~\micron$.
In the compact-{sphere} model, a cloud deck that is gray in visible is produced no matter how small the monomers at the cloud base are,  because they always grow to $\ga 1~{\rm \mu m}$ in size through coagulation as shown in Section \ref{sec:vertical} (see also \citealt{Ohno&Okuzumi18}).
% On the other hand, the compact-sphere clouds do not produce the spectral slope in visible even if very small monomer size (i.e., high number density) is assumed.

The transmission spectrum for fluffy-aggregate clouds exhibit a considerably different shape from that for compact-sphere clouds. 
Since the fluffy-aggregate cloud is lofted to much higher altitude, the absorption features in the spectra are largely obscured as compared to the cases of the compact-sphere clouds {except for the case of $r_{\rm mon}=0.01~{\rm \mu m}$ in which the effect that decreases the cloud extinction efficiency is important (Section \ref{sec:cloud-top}).} 
Furthermore, the fluffy-aggregate clouds produce a spectral slope at $\lambda \la 2~\micron$, in particular when the monomers are small.   
%\edit1{\it I don't see this: The spectra significantly obscure the absorption feature compared to the cases of the compact-sphere clouds because the fluffy-aggregate cloud is lofted to much higher altitude as shown in Figure \ref{fig:vertical}.}
% In contrast to the compact-sphere cloud that makes a gray cloud deck, the fluffy-aggregate cloud tends to produce the slope-like structure in the wide range of wavelength from visible to near-infrared.
The spectrum for $r_{\rm mon}=1~{\rm \mu m}$ is nearly identical between the fluffy-aggregate and compact-{sphere} models because the monomers satisfy $\lambda > 2\pi r_{\rm mon}$ at near-infrared wavelengths. 
The spectral slope for $r_{\rm mon}=0.1$ and $0.01~{\rm \mu m}$ is well characterized by $\mathcal{S}({\alpha}=-2)$, originated by the wavelength dependence of the scattering opacity for $2\pi r_{\rm mon}<\lambda<2\pi r_{\rm agg}$ (see Section \ref{sec:kappa_agg}). 

{ 
Since the spectral slope with $\mathcal{S}({\alpha}=-2)$ originates from the scattering property of aggregates, it could { potentially be} used as an observational signature {for CPAs } {when} the atmospheric scale height $H$ is well constrained.
We find that the slope with $\mathcal{S}({\alpha}=-2)$ also emerges for many other materials that may build up mineral clouds on exoplanets (Appendix \ref{appendix1}).
However, caution should be taken regarding this interpretation because $\mathcal{S}({\alpha}=-2)$ may also be caused by the combination of small and large compact spheres.
}
%\citep[e.g.,][]{Macdonald&Madhusudhan17} might offer the clues to the presence of fluffy aggregates in exoplanetary atmospheres if pressure scale height $H$ is constrained by other information, such as the depth of molecular signature. %\edit1{\it but isn't there a degeneracy between $H$ and $n$?} %\textcolor{blue}{\it If one can only see the spectral slope, there is a degeneracy between $H$ and $n$. But one can constrain $H$ using other signature, for example, the depth of absorption feature of some molecules. Then, it would be possible to constrain the spectral index if we have sufficient information, which maybe associated to the range of observable wavelength.} 

Although the fluffy aggregates can largely obscure the molecular features in visible to near-infrared, they are optically too thin to hide the features at longer wavelengths ($\lambda\ga2~{\rm \mu m}$), as can be seen in Figure \ref{fig:spectrum_met}.
% Figure \ref{fig:spectrum_met} shows that the spectra with fluffy-aggregate clouds are nearly superposed on the spectra with cloud-free atmosphere at $\lambda > 3$--$5~{\rm \mu m}$.
% \edit1{This} implies that even if the spectral feature is obscured by the fuffy-aggregate cloud in near-infrared, 
% This implies that if the high-altitude \edit1{clouds of super-Earths consist} of fluffy aggregates, future observations with JWST and ARIEL, which will probe the wavelength range of $\lambda \ga 2~{\rm \mu m}$, could detect molecular features in super-Earths that look cloudy in visible and near-infrared.
This implies that future transmission spectroscopy at $\lambda\ga 2~\micron$ with JWST and ARIEL could detect molecular features in super-Earths that look cloudy in visible and near-infrared.
%\edit1{{\it I don't understand the following sentence. Is it relevant to what is discussed above?}: Further investigations with observational \edit1{noise} model \citep[e.g.,][]{Batalha+17} and retrieval models \citep[e.g.,][]{Macdonald&Madhusudhan17,Neila+18} would quantify the influences of fluffy-aggregate clouds on future observations, which we defer for future studies. {\it A comment on English: a cloud in an exoplanet can affect observed transmission spectra but cannot affect observations themselves. A cloud on the Earth can affect observations.}}

The transmission spectrum from the fluffy-aggregate model also substantially depends on the atmospheric metallicity. 
Figure \ref{fig:spectrum_met} shows the transmission spectra from the fluffy-aggregate model for various atmospheric metallicities, where $R_{\rm 0}=2.25R_{\rm Earth}$ is assumed for every case.
%Figure \ref{fig:spectrum_met} shows how the transmission spectra from the fractal aggregate model depends on atmospheric metallicity. %\it Let's focus on cloudy cases.}
% For spectra with cloud-free atmospheres, the absorption feature is enhanced from the metallicity of $1\times$ solar to $10\times$ solar, while the spectral features are weaken as the metallicity increases for $\ga 100\times$ solar.
% The enhancement of spectral features are simply caused by the incresae of mixing ratio of gas molecules.
% For $\ga 100\times$ solar metallicity, the pressure scale height $H$ moderately decreases with increasing the metallicity because the metal mass fraction exceeds $0.5$ around the metallicity of $100\times$ solar.
% Since the transit radius decreases with decreasing $H$ \citep[e.g.,][]{Heng&Kitzmann17}, the spectral features are weak in the planets with high atmospheric metallicity.
% As seen in Figure \ref{fig:spectrum_met}, the shape of transmission spectra with the fluffy-aggregate cloud also depend on the atmospheric metallicity.
%In principle, the higher \edit1{the} metallicity is, the weaker \edit1{the} molecular features are, 
%This is due to the increase of cloud mass mixing ratio accompanied with the atmospheric metallicity (see also Figure \ref{fig:vertical}) that makes the upper atmosphere optically thick.
%In all cases, the sub-micron monomer size leads to the appearance of spectral slope except the case of $1000\times$ solar metallicity.  
One can see that the higher the atmospheric metallicity is, the flatter the spectral slope is. This is because the gradient of spectral slope is proportional to the pressure scale height $H$ (see Equation \ref{eq:slope}), which decreases with increasing the atmospheric metallicity. The effect is notable for $\ga100\times$ solar metallicity, and the spectral slope is almost flat for $\ga 1000\times$ solar metallicity.

%due to the scattering by fluffy aggregates is more gentle for higher metallicity, especially for $>100\times$ solar.

%The slope eventually approaches a flat line for $1000\times$ solar metallicity.
%This trend is again caused by the effect of scale height.
%The gradient of spectral slope is given by \citep[e.g.,][]{Line&Parmentier16},
%\begin{equation}\label{eq:slope}
%\frac{dD(\lambda)}{d\log{\lambda}}=\frac{2\pi R_{\rm p}H}{\pi R_{\rm *}^2}\alpha,
%\end{equation}
%\edit1{\it Eq. 40 should be derived in the previous section. This explains everything about the scattering slope, right? We could shorten the text if we show Eq. 40 earlier.} 
%Equation \eqref{eq:slope} indicates that the large pressure scale height and $|\alpha|$ produce a steep spectral slope, and vice versa for the small scale height and $|\alpha|$.
%This is why the spectral slope approaches nearly flat line as the metallicity increases.

%%%%%%%%%%%%%%%%%%%%%%%%%%%%%%%%%%%%%%%%%%%%%%%
\subsection{Comparison with Observations of GJ1214b}\label{sec:GJ1214b}
%%%%%%%%%%%%%%%%%%%%%%%%%%%%%%%%%%%%%%%%%%%%%%%
\begin{figure*}[t]
\centering
\includegraphics[clip,width=\hsize]{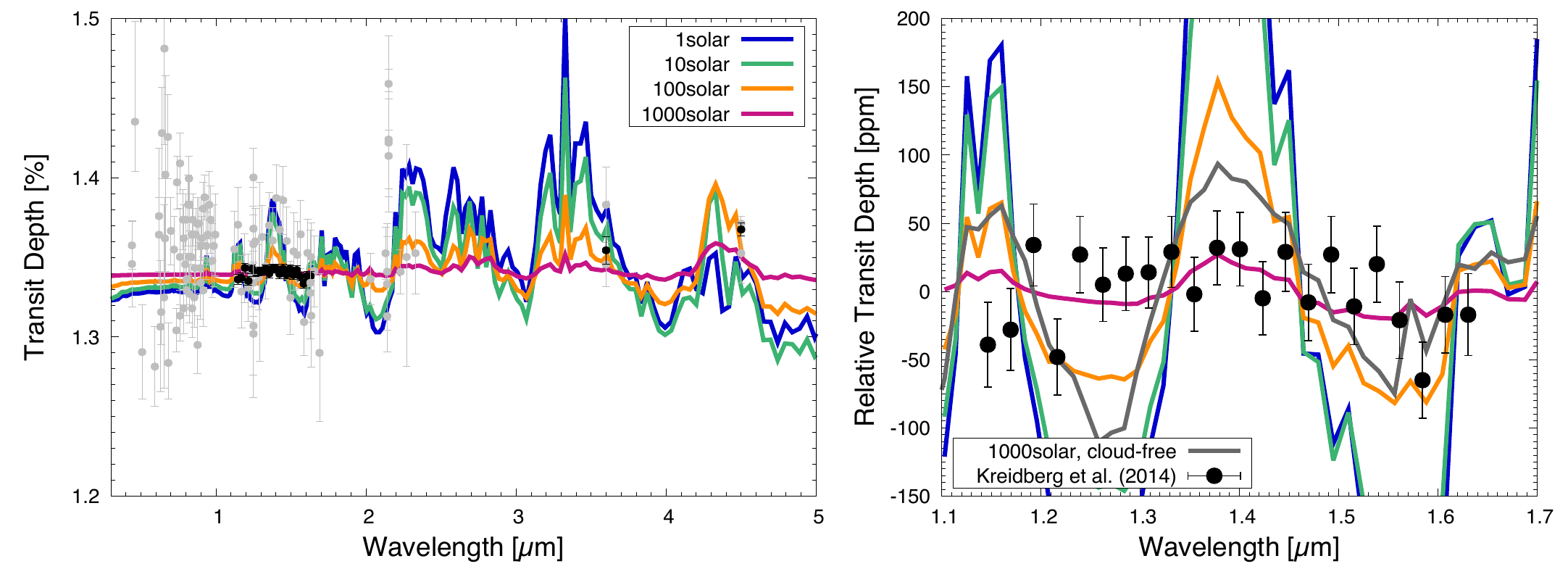}
\caption{Synthetic transmission spectra of GJ1214 b (colored lines) compared with the observational spectrum to date (black and gray points). The left panel shows all observed transit depth ranging from $0.3$ to $5~{\rm \mu m}$ and {the best-fit spectra for comparisons with data of HST/WFC3 \citep{Kreidberg+14} and Spitzer/IRAC \citep{Gillon+14}.} The right panel shows the comparisons with the data points of only HST/WFC3. The horizontal axis are wavelength, and the vertical axises are transit depth. The blue, green, orange, and red lines show the spectra with the metallicity of $1\times$ solar and with $r_{\rm mon}=0.3~{\rm \mu m}$ (reduced chi-square is $\chi_{\rm red}^{2}=57.13$), $10\times$ solar and $r_{\rm mon}=0.3~{\rm \mu m}$ ($\chi_{\rm red}^{2}=37.13$), $100\times$ solar and $r_{\rm mon}=0.3~{\rm \mu m}$ ($\chi_{\rm red}^{2}=6.62$), and $1000\times$ solar and $r_{\rm mon}=0.3~{\rm \mu m}$ ($\chi_{\rm red}^{2}=2.03$), respectively.
The black line in the right panel also shows the best-fit spectrum for cloud-free atmosphere with the $1000\times$ solar metallicity.
The spectral resolution is binned down to $\lambda/\Delta \lambda \approx 100$ for clarity. The gray dots exhibit currently available observational data \citep{Bean+11,Croll+11,Desert+11,deMooij+12,Berta+12,Murgas+12,Colon&Gaidos13,Narita+13a,Narita+13b,Fraine+13,Rackham+17}. Specifically, the black dots indicate the data from the observations by the HST/WFC3 \citep{Kreidberg+14} and the Spitzer/IRAC \citep{Gillon+14}.}\label{fig:transmission}
\end{figure*}
%%%%%%%%%%%%%%%%%%%%%%%%%%%%%%%%%%%%%%%%%%%%%%%

Here, we compare our synthetic transmission spectra with the observed transmission spectrum of GJ1214b. %\edit1{{\it The following introduction should be presented earlier, not here:} A number of observations of transmission spectra have been carried out for this planet \citep[e.g.,][]{Bean+11,Narita+13b,Fraine+13,Kreidberg+14,Rackham+17}.
%Nevertheless, the atmospheric composition is still highly uncertain because of the extremely featureless spectrum that is possibly caused by the clouds in very high altitude such as $\sim 0.01~{\rm mbar}$ \citep{Kreidberg+14}.}
We calculate the cloud profiles as well as the synthetic spectra for the atmospheric metallicities of $1$--$1000\times$ solar abundances and monomer sizes of $0.01$--$1~{\rm \mu m}$.
We also vary the reference radius $R_{\rm 0}$ from $2$ to $3R_{\rm Earth}$ to be consistent with the observed planet radius. %\edit1{\it Why didn't you adopt $3R_{\rm Earth}$ in the previous subsection?.} 
The relative goodness-of-fit for each model is quantified by the reduced chi-square $\chi_{\rm red}^2$. The model freedom is the number of data points minus three, the number of the fitting parameters (atmospheric metallicity, monomer size, and reference radius). For instance, \citet{Morley+15} assumed that an acceptable model for GJ1214b produces $\chi_{\rm red}^2<1.14$ if only data points from the HST/WFC3 observations \citep{Kreidberg+14} are used.

\begin{figure*}[t]
\includegraphics[clip,width=\hsize]{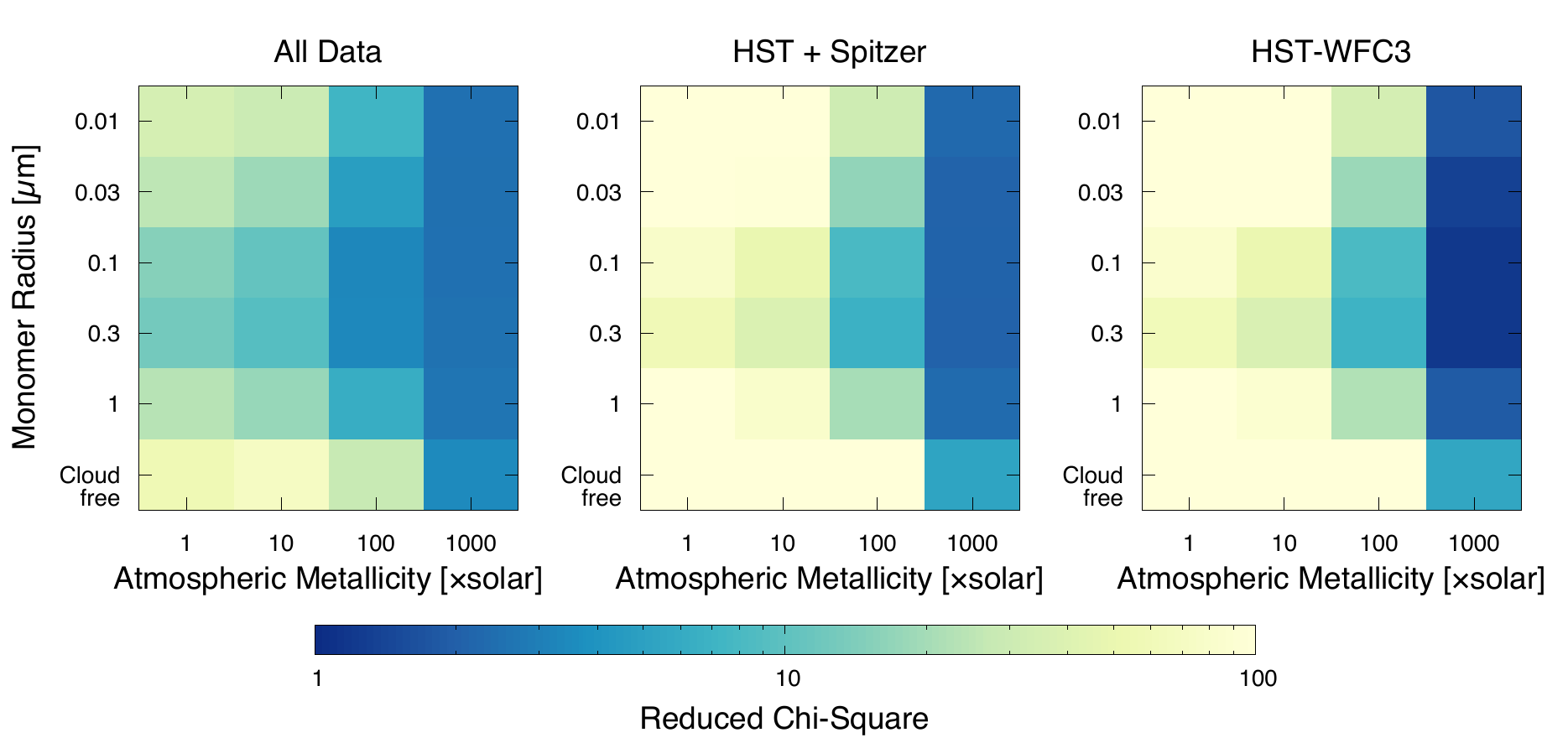}
\caption{ {Reduced chi-squared values for the synthetic transmission spectra of GJ1214b as a function of the monomer radius and atmospheric metallicity.
% The colorscale indicates the reduced chi-square for $3$ fitting parameters (atmospheric metallicity, monomer radius, and reference planet radius). 
The left panel shows the chi-squred values obtained by fitting models to all observational data. The middle panel shows the results from the analysis that only uses the data of HST/WFC3 \citep{Kreidberg+14} and Spitzer/IRAC \citep{Gillon+14}. The right panel is from the analysis that only uses the  data of HST/WFC3.}
%The left panel shows the results of fitting for all observational data, while the right panel shows those for only the data of \citet{Kreidberg+14}. The bottom rows labeled by ``Free'' exhibit the reduced chi-square for cloud-free atmospheres.
}\label{fig:chi}
\end{figure*}

% We find that the model with a higher atmospheric metallicity yields a better match \edit1{\sout{to}} the observational data.
{The left panel of} Figure \ref{fig:transmission} shows the best-fit transmission spectra for the metallicity of $1$, $10$, $100$, and $1000\times$ solar abundance{,} compared with { {the} observational data {for GJ1214b from} HST/WFC3 \citep{Kreidberg+14} and Spitzer/IRAC \citep{Gillon+14}. {All available observational data are also denoted as gray dots in the left panel of Figure \ref{fig:transmission}.}
{The} right panel shows the best-fit spectra only for HST/WFC3. 
}
%\edit1{\it Describe more details  about model fitting, not here but in the last paragraph. What are the model parameters and how the best-fit model is selected? Maybe the description on the reduced chi-square fitting presented below should also be moved}.
%One can  see that the data in visible is largely scattered each other, and any models cannot reproduce all of them simultaneously.
%The large scattering of observational data points might be due to the stellar heterogeneity effects \citep{Rackham+17,Rackham+18}, which is significant in visible.
%Therefore, we also perform the comparisons with the data only from HST/WFC3 observation of \citet{Kreidberg+14} as done by \citet{Morley+15} and \citet{Charnay+15} (right panel of Figure \ref{fig:transmission}).
For all data points (left panel), the smallest reduced chi-square for the atmospheric metallicities of $1$, $10$, $100$, and $1000\times$ solar abundance are $\chi_{\rm red}^{2}=12.33$, $9.04$, $3.28$, and $2.41$, respectively.
{ {However, these $\chi_{\rm red}^{2}$ values are significantly affected by the large scatter in the data  at visible wavelengths}.
If we only focus on the data point of HST/WFC3 \citep{Kreidberg+14} and Spitzer/IRAC \citep{Gillon+14}, which are less scattered than the visible data, we obtain the reduced {chi-squared} of $\chi_{\rm red}^{2}=57.13$, $37.13$, $6.62$, and $2.03$ for the metallicities of $1$, $10$, $100$, and $1000\times$ solar abundance, respectively.
For a comparison with the HST data only (right panel), the reduced {chi-squared values are} $\chi_{\rm red}^{2}=60.23$,$36.50$, $6.78$, and $1.16$ for atmospheric metallicities of $1$, $10$, $100$, and $1000\times$ solar abundance, respectively.
Overall, {a} higher atmospheric metallicity leads to {a} smaller reduced {chi-squared value}. 
We also find that the presence of the fluffy-aggregate cloud appreciably improves the goodness-of-fit of the model as compared to the cloud-free case.
}
{For the comparison with the HST data as an example}, the cloud-free atmosphere with $1000\times$ solar metallicity yields $\chi_{\rm red}^2=5.44$ (the black line in the right panel of Figure \ref{fig:transmission}){, whereas the model with the fluffy-aggregate clouds yields $\chi_{\rm red}^2=1.16$.} 
The reduced chi-square $\chi_{\rm red}^2$ for each parameter set is summarized in Figure \ref{fig:chi}.

Our results show that the model with a higher atmospheric metallicity yields a better match to the observational data.
% In contrast, the models with high-metallicity atmospheres ($100$ and $1000\times$ solar) can overcome the obstacles for low-metallicity atmospheres.
The high-metallicity atmospheres supply sufficient KCl condensates, and the produced clouds can obscure the molecular features if monomer size is sufficiently small, namely ${\la}1~{\rm \mu m}$. 
Indeed, the molecular absorption at around $\lambda=1.4~{\rm \mu m}$, noticeable in cloud-free atmospheres even with $1000\times$ solar metallicity, is significantly weakened by the cloud opacity (right panel of Figure \ref{fig:transmission}).
The spectral slope is also closer to the observed flat spectrum because of the relatively small scale height.
%The high-metallicity models also reasonably match to the observations of the Spitzer IRAC ($\lambda=3.6$ and $4.5~{\rm \mu m}$ in left panel).
Notably, the model with $1000\times$ solar metallicity yields $\chi_{\rm \nu}^2=1.16$ for the comparisons with \citet{Kreidberg+14}, which is comparable to the $\chi_{\rm red}^2$ obtained by \citet{Gao&Benneke18} who assumed the eddy diffusion coefficient much larger than that predicted by 3D GCM \citep{Charnay+15}.
Our results suggest that it would be able to explain the observed spectra of GJ1214b in the range of $K_{\rm z}$ predicted by the GCM, if the mineral cloud consist of fluffy aggregates.
We emphasize that { {as the metallicity is increased, the resulting} synthetic spectrum better matches the transit depth at mid-infrared wavelength{s}, especially at $4.5~{\rm \mu m}$, observed by the Spitzer/IRAC (see the left panel of Figure \ref{fig:transmission}).} 
%the model with high-metallicity atmospheres also reasonably matches the transit depth at mid-infrared wavelength, especially at $\lambda=4.5~{\rm \mu m}$, observed by the Spitzer IRAC.
This is thanks to the absorption of $\rm CO_2$ {whose abundance increases with increasing the atmospheric metallicity} \citep{Moses+13}.

%%%%%%%%%%%%%%%%%%%%%%%%%%%%
%%%%%%%%%%%%%%%%%%%%%%%%%%%%

%We find that the high-metallicity model is preferred to explain the observational data as compared to the low-metallicity model.
%As seen in Figure \ref{fig:chi}, the models with low atmospheric metallicities ($1$ and $10\times$ solar metallicity) are clearly inconsistent with the observational data, even if the monomer size is very small.
There are two reasons why the low-metallicity models ($1$ and $10\times$ solar) fail to match the observations.
The first is an insufficient cloud abundance: the mixing ratio of KCl in the low-metallicity atmosphere is too low to produce sufficiently opaque clouds.
The second reason is that, more importantly, the spectral slope caused by the aggregate opacity is too steep to match the flat spectrum observed by the HST/WFC3 \citep{Kreidberg+14} because of the large scale height (see Section \ref{sec:spectrum}).
Therefore, our fluffy-aggregate cloud model still requires a small atmospheric scale height to explain the flat spectrum of GJ1214b, which is achieved by the high-metallicity atmosphere.  

Although the fluffy-aggregate clouds potentially explain the featureless spectrum of GJ1214b, we note that the CPAs need to be constituted by monomers with $r\la 1~{\rm \mu m}$ (see Figure \ref{fig:chi}).
%Unfortunately, the monomer size is highly uncertain for exoplanetary atmospheres.
%\edit1{\it The following sentence is grammatically incorrect. Try shorten the sentence} 
The monomer size is presumably controlled by the formation of condensation nuclei, namely the nucleation, and subsequent condensation growth that keeps a spherical shape \citep{Lavvas+11}.
If one adopts the classical nucleation theory, the homogeneous nucleation followed by condensation yields KCl particles with the effective size of ${\sim}{10}~{\rm \mu m}$ \citep{Gao&Benneke18}.
This is substantially larger than the required monomer size.
This could suggest that classical nucleation theory underestimates the nucleation rate of $\rm KCl$, 
because a larger number of condensation nuclei generally leads to a smaller monomer size \citep{Gao+18,Ohno&Okuzumi18}. In fact, \citet[][]{Lee+18} reports that classical nucleation theory underestimates the nucleation rate of $\rm TiO_{2}$.
Alternatively, a number of stable, small nuclei  could be produced by the heterogeneous nucleation of $\rm ZnS$ onto $\rm KCl$ \citep{Gao&Benneke18}, although its nucleation rate depends on the desorption energy of $\rm ZnS$ that is currently unknown.
%\it Why can the additional condensation of ZnS onto KCl lead to smaller monomers? Can ZnS stop the homogeneous nucleation of KCl?} 
Laboratory studies of nucleation and condensation would be important to predict the monomer size in exoplanetary atmospheres, which in turn test the scenario of the fluffy-aggregate cloud for GJ1214b.

%The monomer size will be determined as a size for which condensation and coagulation timescales are equal, since the condensation produces sphere-like particles \citep{Lavvas+11}.

%The monomer size has been highly uncertain for exoplanetary atmospheres.

%\citep[][]{Desert+11,Fraine+13,Gillon+14}. 
%We note that, despite the model difference, the high-metallicity atmosphere was also favored to explain the observations of GJ1214b in previous studies \citep[][]{Morley+15,Charnay+15b,Gao&Benneke18}.

%, which might offer the hints to constrain the properties of building-block of planets 
%from the atmospheric metallicity.
%Further studies linking the atmospheric composition to the planet formation processes \citep[e.g,][]{Mordasini+16,Cridland+16,Cridland+17} would be helpful to explore the nature of exoplanets.

%%%%%%%%%%%%%%%%%%%%%%%%%%%%%%%%%
\section{Discussion}\label{sec:discussion}
%%%%%%%%%%%%%%%%%%%%%%%%%%%%%%%%%
\subsection{Model Caveats}
In this study, we have adopted simplified porosity and microphysical model.
The models are useful to clarify the effects of the porosity evolution, but involves some caveats because of its simplicity.
In what follows, we state the caveats of our model and discuss their possible impacts on the results.

\subsubsection{Validity of $D_{\rm f}=2$ for Other Size Distributions}\label{sec:Df_test}
The most strong assumption of our porosity model may be the fractal dimension of $2$ for the fractal growth (Section \ref{sec:fractal}).
We have adopted this assumption since our cloud microphysical model assumes the narrowly peaked size distribution, for which the equal-sized collision is a dominant growth process.
However, the cloud particles could have different shape of size distributions \citep{Powell+18,Gao&Benneke18}, and the monomer-aggregate collision might be dominant. 
In that case, CPAs grow into more spherical shapes (e.g., $D_{\rm f}\approx 3$), and the cloud vertical extent would be small as compared to the case of $D_{\rm f}=2$.

Here, we test the validity of the assumption $D_{\rm f}=2$ for various size distributions. 
We introduce a mass-weighted collision rate onto a particle with mass $m_{\rm t}$ ($m\leq m_{\rm t}$), defined as \citep{Okuzumi09}
\begin{equation}\label{eq:C_coll}
    C_{\rm m_{\rm t}}(m)=\frac{mK(m_{\rm t},m)f(m)}{\int_{\rm 0}^{m_{\rm t}} m'K(m_{\rm t},m')f(m')dm'},
\end{equation}
where $K(m,m')$ is the collision kernel between particles with masses $m$ and $m'$, and $f(m)dm$ is the number density of particles with masses between $m$ and $m+dm$.
Equation \eqref{eq:C_coll} measures the contributions of aggregates with masses of $m$ on the growth of aggregate with mass of $m_{\rm t}$.
We assume that the cloud particles obey the Hansen size distribution \citep{Hansen71}, described as
\begin{equation}
    f(r)\equiv \frac{dn(r)}{dr}\propto r^{(1-3b)/b}\exp{\left(-\frac{r}{ab}\right)},
\end{equation}
where $a$ is the mean effective radius and $b$ is the effective variance.
The shape of the size distribution is controlled by the effective variance $b$; for example, $b<0.5$ yields log-normal-like distributions, while $b>0.5$ yields power-low-like distributions.
%and can be fitted by $b=0.1$--$0.2$ for terrestrial water clouds and $b>0.5$ for cloudy brown dwarfs \citep{Hansen71,Hiranaka+16}.

\begin{figure}[t]
\includegraphics[clip,width=\hsize]{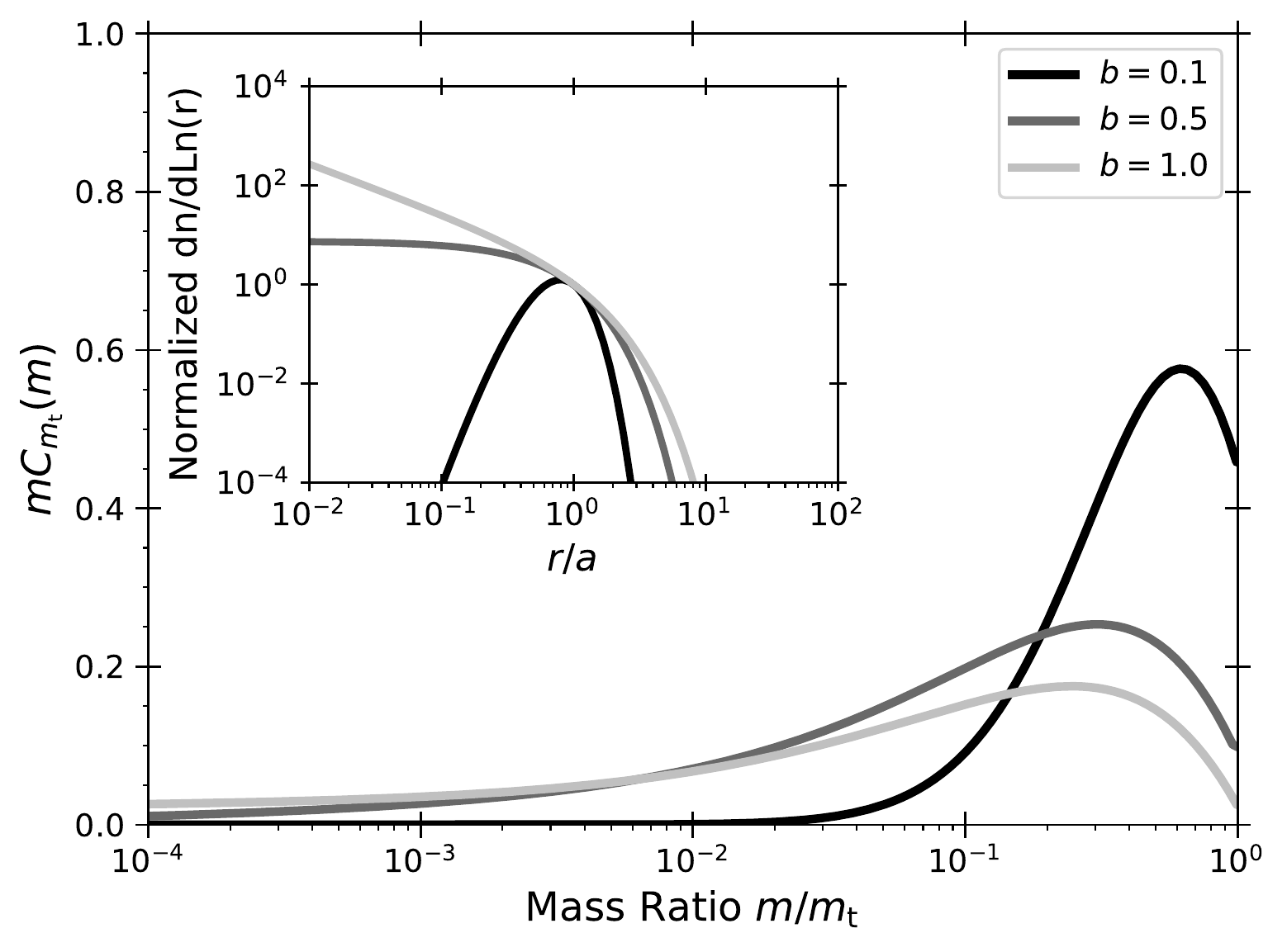}
\caption{Normalized mass-weighted collision rate between particles with masses $m_{\rm t}$ and $m$. The black, gray, and silver lines show the collision rate for the Hansen size distributions with $b=0.1$, $0.5$, and $1.0$, respectively. The corresponding size distributions normalized by $af(a)$ are also shown in the inner panel.}\label{fig:C_coll}
%The solid and dashed lines show the evolution tracks without and with collisional compression, respectively.}
\end{figure}

Figure \ref{fig:C_coll} shows the mass-weighted collision rate and size distributions for $b=0.1$, $0.5$, and $1.0$. 
We assume $a=1~{\rm \mu m}$ and $m_{\rm t}$ calculated from the mass-weighted particle size:
\begin{equation}
    m_{\rm t}=\frac{4\pi \rho_{\rm p}}{3}\left( \frac{\int_{\rm 0}^{\rm \infty} rmf(r)dr}{\int_{\rm 0}^{\rm \infty} mf(r)dr} \right)^3=\frac{4\pi \rho_{\rm p}}{3}a^3(1+b)^3.
\end{equation}
We use the collision kernel described in Chapter 15 of \citet[][]{Jacobson05} assuming a constant particle density.
%assuming atmospheric pressure of $0.1~{\rm bar}$ and $T=1000~{\rm K}$.
Figure \ref{fig:C_coll} demonstrates that the growth is largely contributed by the collisions of particles with masses of $m/m_{\rm t}\sim 0.01$--$1$.
According to \citet{Okuzumi+09}, collisions with mass ratio of $0.01$--$1$ lead to the fractal dimension of $D_{\rm f}\sim 1.9$--$2.1$ (see their Figure 6), which is almost the same as $D_{\rm f}=2$ assumed in this study.
Thus, the assumption of $D_{\rm f}=2$ may be reasonable for various shapes of size distributions.
However, it should be noted that the size distribution of the CPAs has been unknown to date.
We will examine how the size and porosity distributions of CPAs evolve in exoplanetary atmospheres in our forthcoming paper.
%However, the size distributions of the CPAs have not been examined yet, and we will investigate it in our forthcoming paper.

\subsubsection{Limitation of {the} Compression Model}
Here we state several limitations of  a compression model adopted in Section \ref{sec:porosity_recipe}. 
%There are several limitations in our compression model.
First, the relation between mass and size of the collisionally compressed aggregates (Equation \ref{eq:Wada}) was derived for collisions between two equal-mass aggregates with $D_{\rm f}\approx 2$ \citep{Wada+08}. 
For different-mass collisions, the degree of compression is evaluated from the comparison of the impact energy with work done by dynamic compression strength \citep{Suyama+12}. 
we also note that the head-on-collision is assumed here, but offset collisions could induce the elongation of aggregates, further hinders the compression \citep{Paszun&Dominik09}. 
Second, the static compression strength used for the gas-drag compression (Equation \ref{eq:P_str}) was derived for an aggregate whose internal structure is characterized by $D_{\rm f}\approx2$ \citep{Kataoka+13a}.
{The compression strength for different $D_{\rm f}$ was recently proposed by \citet{Arakawa+19} from a semi-analytical argument.
Although the verification with numerical experiments remain to be carried out, their formula is potentially applicable to our compression model.}
%The compression strength for different internal sructure remains to be studied.
Further numerical experiments will be helpful to extend the compression model to more universal cases.

%and not tested for that with arbitrary $D_{\rm f}$. 
%The static compression strength has not been tested for other structure, such as a chain-like aggregate ($D_{\rm f}\approx 1.5$). 
%Nevertheless, we do not attempt to further improve the compression model since the assumption of $D_{\rm f}\approx2$ is likely valid in many cases (Section \ref{sec:Df_test}), and our current microphysical model does not handle the collisional growth between different-mass aggregates.
%Further numerical experiments will be needed to extend our model to general aggregates, which is beyond the scope of this study.}

\subsubsection{Simplified Nucleation and Condensation}
%In this study, to mimic the nucleation followed by condensation growth, namely the monomer formation, we have set the lower boundary condition assuming that saturated vapor instantaneously incorporated into the condensation nuclei. 
In this study, we have assumed that saturated vapor is instantaneously incorporated into the condensation nuclei at the cloud base. 
This assumption would be reasonable since the condensation timescale is much shorter than the vertical mixing timescale near at the cloud base \citep{Ohno&Okuzumi18,Powell+18,Gao&Benneke18}.
Additional condensation could transform the CPAs to sphere-like particles if the surface growth rate via condensation dominates over the coagulation rate \citep{Lavvas+11}.
However, the effect is presumably insignificant for KCl clouds because other condensing species, such as $\rm {Na}_2S$ and MnS, have the cloud bases at deeper atmospheres and are likely depleted at the KCl cloud formation region \citep[e.g.,][]{Mbarek&Kempton16}.
{ZnS is an only species whose cloud base is placed near at the KCl cloud base \citep[e.g.,][]{Morley+12}.
But, we expect that CPAs are still present as aggregates even if ZnS condensation takes place.
This is because the abundance of ZnS is 2--3 times lower than KCl and likely insufficient to fill all pores.
}
%are likely depleted through the ``rain-out'' effects in the deep atmosphere \citep[e.g.,][]{Mbarek&Kempton16}.
%Additional condensation might transform the CPAs to graupel-like particles.

We have also assumed that the nucleation followed by condensation, namely the monomer formation, occurs right at the cloud base.
This would be true if the condensation nuclei are supplied from deep atmospheres, as argued in \citet{Lee+18}.
On the other hand, the monomer formation could occur above the cloud base in the context of homogeneous nucleation that needs significant supersaturation to set in \citep[e.g.,][]{Helling&Fomins13}.
The monomers formed in upper atmospheres might increase the $D_{\rm f}$ of CPAs through different-size collisions.
We expect that the resulting $D_{\rm f}$ is still close to $2$, as discussed in Section \ref{sec:Df_test}, though a microphysical model solving size distributions will be needed to verify it.

%\subsubsection{\bf Possibility of Molten KCl Clouds}

%%%%%%%%%%%%%%%%%%%%%%%%%%%%
\subsection{Comparison with Other Porosity Models}
\begin{figure}[t]
\includegraphics[clip,width=\hsize]{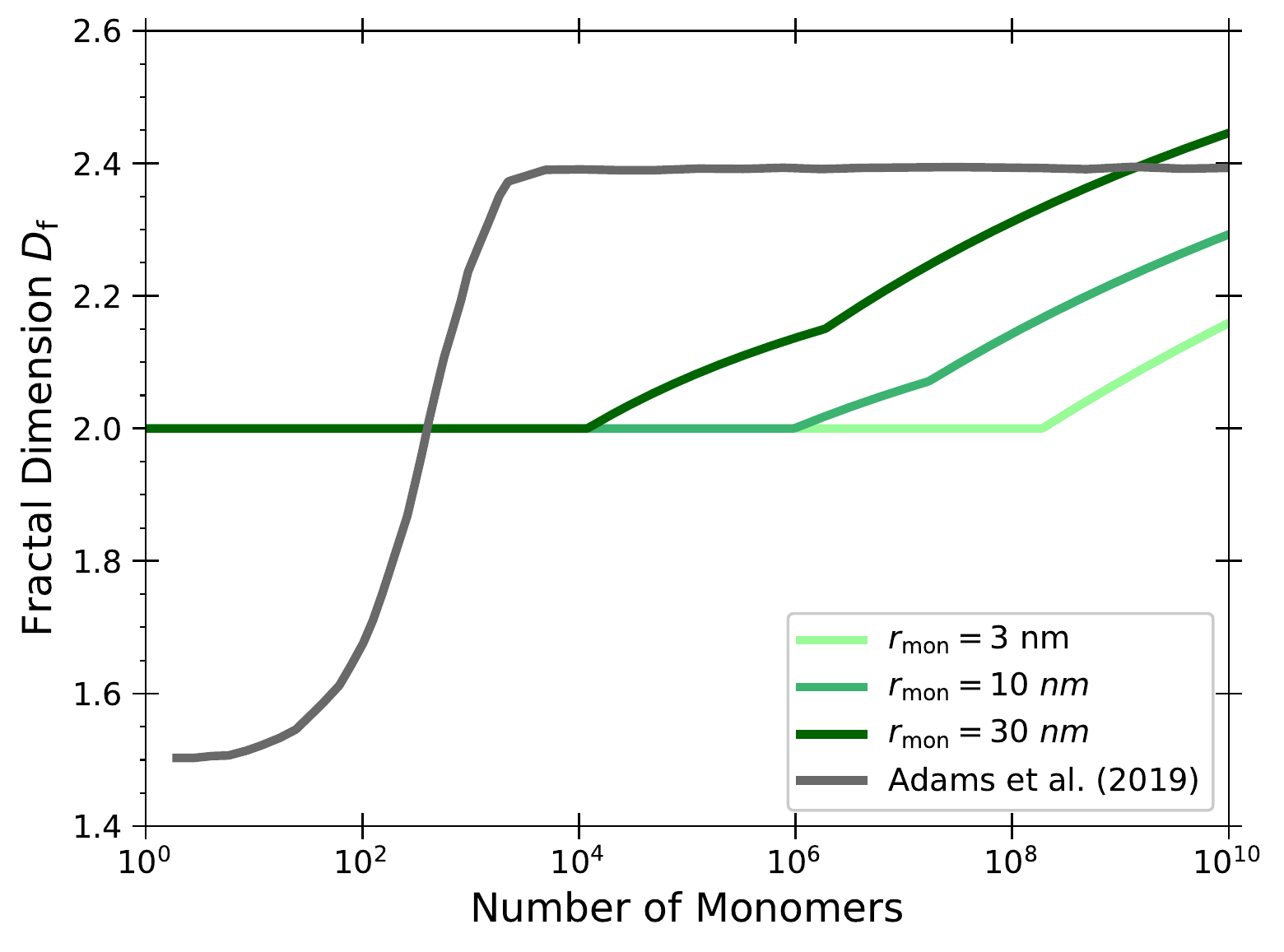}
\caption{Comparison of our porosity model with that used in \citet{Adams+19}. The vertical and horizontal axes show the fractal dimension $D_{\rm f}$ and number of monomers $N_{\rm mon}$. Different colored lines exhibit the evolution track of $D_{\rm f}$ for different monomer size, and the gray line shows the track assumed in \citet[][]{Adams+19}. We assume $P=0.01~{\rm mbar}$ to evaluate the collision velocity.}\label{fig:Df_list}
%The solid and dashed lines show the evolution tracks without and with collisional compression, respectively.}
\end{figure}
%Recently, \citet{Adams+19} studied the effects of porosity evolution on the photochemical haze formation and transmission spectra adopting a porosity evolution model different from our model. %using a microphysical model.
%They suggested that the opacity of haze-aggregates is gray and largely flatten transmission spectra, which is in contrast to our results in which the cloud particle aggregates tend to produce the spectral slope.
%The discrepancy is presumably caused by the different composition of aerosols and different model of the porosity evolution.
%We discuss differences between 

Some previous studies of haze microphysics adopted a porosity model different from ours \citep{Wolf&Toon10,Adams+19}.
The porosity model adopted in the haze models assumes that the fractal dimension approaches $D_{\rm f}\approx 2.4$ as the number of monomers increase.
The assumed fractal dimension is comparative to $D_{\rm f}=2.5$ that was observed for an aggregates with maximal compression via high-energy collisions in the numerical experiments \citep{Wada+08,Suyama+08}.
However, the threshold at which compression sets in is considerably different from our model.
Figure \ref{fig:Df_list} shows the fractal dimension as a function of the number of monomers in \citet{Adams+19} and our model, where we calculate $D_{\rm f}$ from Equations \eqref{eq:mass-relation} and \eqref{eq:Df}:
\begin{equation}
    D_{\rm f}=3\left( 1-\frac{\log{\phi_{\rm eq}}}{\log{N_{\rm mon}}}\right)^{-1}.
\end{equation}
For comparisons, we use the surface energy of tholine $\gamma=0.0709~{\rm J~m^{-2}}$ \citep{Yu+17} and material density $\rho_{\rm p}=1~{\rm g~{cm}^{-3}}$. 
In the model of \citet{Adams+19}, the fractal dimension increases to $\approx2.4$ at $N_{\rm mon}\ga{10}^{3}$, while our model predicts that the compression sets in $N_{\rm mon}>{10}^4$--${10}^8$, depending on the monomer size.
Thus, the aggregate hazes in previous studies were assumed to be compressed much easier than our prediction.
This is presumably a reason why aggregate hazes in \citet{Adams+19} tend to produce flat spectra rather than those with spectral slopes.

The easily compressed aggregates in previous studies were speculated from the laboratory study of soot formation in a flame.
In the experiments, it was observed that the soot-aggregates are restructured by the Coulomb interaction between oppositely charged parts \citep{Onischuk+03}.
However, one should take a caution about the compression due to the Coulomb interaction because the charge states of aerosols in exoplanetary atmospheres are poorly known.
Investigating the aerosol charge processes \citep[e.g.,][]{Helling+11b,Helling+11a} might help to evaluate if the restructuring due to Coulomb interaction is possible.

\subsection{ Implications for Spectral Slopes of Hot-Jupiters}
{
%Although we have focused on warm super-Earths, hot-Jupiters would be good targets to search the fluffy-aggregate clouds.
%Because their atmospheres should be mainly composed of hydrogen and helium, the atmospheric scale height $H$ and wavelength dependence of cloud opacity $\alpha$ could be constrained easier than that for super-Earths. 
The presence of mineral clouds has {also been} suggested for a number of hot-Jupiters \citep[e.g.,][]{Sing+16,Barstow+17}.
{A recent} retrieval study {by} \citet{Pinhas+19} suggested that the  hot-Jupiters {whose transmission spectra were provided by \citet{Sing+16} typically exhibit  transmission spectral slopes of} $\alpha \la -5$.
This is {considerably} steeper than the slope originated from the aggregate scattering opacity ($\alpha=-2$) and even {steeper than the Rayleigh slope} ($\alpha=-4$). 
Such ``super-Rayleigh'' slope{s} might be produced by the absorption of tiny cloud particles made of sulfide minerals \citep{Pinhas&Madhusudhan17}.
However, we find that CPAs made of sulfide minerals do not produce such steep wavelength dependence (see Figure \ref{fig:appendix1} in Appendix \ref{appendix1}) unless the aggregate is extremely small.
This is because the steep absorption feature of sulfide minerals is largely obscured by the {scattering opacity}.
%This is because the wavelength dependence of the scattering opacity ($\propto \lambda^{-2}$) largely smears out the steep absorption feature in the extinction opacity.
Thus, {it is more likely that the super-Rayleigh slopes of hot-Jupiters are caused by other  physical processes than fluffy-aggregate cloud formation}, such as NUV absorbers like SH \citep{Evans+18}.
Alternatively, the slope potentially implies physical processes that halt the aggregation, leading to a tiny particle size. 
Electrostatic repulsion \citep[e.g.,][]{Okuzumi09} may be promising because the ionization of alkali metals, likely produces charged cloud particles, takes place at hot-Jupiters \citep[e.g.,][]{Batygin&Stevenson10}. 
We will examine this possibility in future studies.
%which produces substantial electrons operating on cloud formation, which we will examine in future studies.

%This looks far from $\alpha=-2$ originated by the aggregate scattering opacity; however, it is important to recall that mineral clouds in hot-Jupiters could be made of condensates different from KCl studied here.
%that induce strong absorption in visible.
%In fact, \citet{Pinhas&Madhusudhan17} showed that the absorption of MnS condensates, likely formed at $T=1100$--$1500~{\rm K}$, can produce the slope-like spectrum with $\alpha{\la}-5$. 
%Since the absorption opacity of an aggregate is the same as the individual monomer (Section \ref{sec:kappa_agg}), some of the steep spectral slopes might be explained by CPAs made of MnS, as produced by tiny MnS particles.
%Our future studies will examine the formation of fluffy-aggregate clouds in hot-Jupiters.
}

\subsection{Implications for High Metallicity Atmospheres on Planetary Formation}
The high-metallicity atmosphere is of interest from the perspective of planetary formation theory.
Our results suggest that, if the flat spectrum of GJ1214b is caused by the condensation clouds, high-metallicity atmosphere ($\ga100\times$ soar) is plausible to explain the observations, as suggested by other studies \citep{Morley+15,Gao&Benneke18}.
{This is in contrast to {some other} super-Earths or exo-Neptunes that likely retain metal-poor ($<100\times$ soar) atmospheres, such as GJ3470b \citep{Benneke+19} and HAT-P-26b \citep{Wakeford+17,Macdonald&Madhusudhan19}.}
On the other hand, metal-rich ($>100\times$ soar) atmospheres have {also been} suggested for some exo-Neptunes, such as GJ436b \citep{Morley+17} and HAT-P-11b \citep{Fraine+14}.
%, as well as for Uranus and Neptune in our solar system \citep{Kreidberg+15}.
{The diversity of the atmospheric metallicity potentially suggests different formation processes of these planets. 
For example, {planets with a low-metallicity atmosphere  may have formed from} large building blocks, such as protoplanets, that less affect atmospheric composition \citep{Fortney+13}.}
{ {{A} high-metallicity atmosphere may suggest that {a substantial metal-enrichment of the atmosphere, potentially caused by the accretion of small planetesimals and/or pebbles \citep[][]{Fortney+13,Lambrechts+14,Venturini+16,Venturini&Helled17}, occurred during the formation of the planet}. }}
%{\it the following sentence duplicates the topic sentence. \sout{What causes the diversity of the atmospheric metallicity would be an interesting topic for planet formation theory.} }} }

The presence of high-metallicity atmospheres poses another interesting question associated to the past formation process: how did the super-Earths avoid to be gas giants?
It has been suggested that the high atmospheric metallicity leads to the runnaway gas accretion even for planets with Earth-masses \citep{Hori&Ikoma11,Venturini+15}.
Thus, the gas accretion must be inhibited in order to form a super-Earth rather than a gas giant. 
One of the scenario is that they were formed in the late stage of protoplanetary disks where the disk gasses were almost dissipated \citep[e.g.,][]{Ikoma&Hori12,Lee+14,Lee&Chiang16}.
Alternatively, the high-metallicity atmospheres may suggest the presence of mechanisms regulating the gas accretion, such as the gap formation and weak viscous accretion of disc gasses \citep{Tanigawa&Ikoma07,Tanigawa&Tanaka16,Ogihara&Hori18}.
Rapid recycling of the atmospheric gas embedded in protoplanetary disc, which is observed in recent hydrodynamical simulations \citep[e.g.,][]{Ormel+15,Lambrechts&Lega17,Kurokawa&Tanigawa18,Kuwahara+19}, also delays the runnaway gas accretion, but it might be difficult to produce the high-metallicity atmosphere unless the disc gas is highly enriched in heavy elements.
The evolution of atmospheric composition after the disk dissipation, like that suggested for solar-system terrestrial planets \citep{Sakuraba+19}, might increase the atmospheric metallicity even if the planet originally possessed a low-metallicity atmosphere.
Further studies linking the formation processes to the atmospheric metallicity would be warranted to explore the past formation processes of super-Earths {with high-metallicity atmospheres}.
\section{Summary}\label{sec:summary}
We have investigated how the porosity of cloud particle aggregates (CPAs) evolve in exoplanetary atmospheres.
Based on the results of numerical experiments investigating the aggregate restructuring, we have constructed a porosity evolution model that takes into account the fractal growth, collisional compression, and the compression caused by gas drag.
Using a cloud microphysical model coupled with the porosity model, we have examined how the porosity evolution influences the cloud vertical distributions and observed transmission spectra of GJ1214b.
Our findings are summarized as follows.

(1) The internal density of CPAs can be much lower than the material density by $1$--$3$ orders of magnitudes (Section \ref{sec:method}), depending on the size of monomers. The gas-drag compression sets in once the CPA becomes larger than $\approx{30}~{\rm \mu m}$ (Equation \ref{eq:r_comp}). The collisional compression is less important than the gas-drag compression in most cases studied here. 
%This is because the mineral clouds are formed at deep atmospheres where the the particle settling velocity is slow.

%The collisional compression occurs only when the monomers are large, or the collision takes place at high altitude. 
    %Alternatively, the compression occurs if the aggregate size exceeds the compression radius, which is $\sim {30}~{\rm \mu m}$ in the case of $D_{\rm f}=2$ (Equation \eqref{eq:r_comp}).

(2) The compression of CPAs hardly occurs during the KCl cloud formation since the particle growth is not sufficient to induce the compression (Section \ref{sec:vertical}). Thus, the porosity evolution in general results in the cloud vertical extent much larger than that of the compact-sphere clouds. Without the compression, the fluffy-aggregate clouds can ascend to the height where the monomer can ascend to (Equation \ref{eq:Ptop}).
%size and vertical extent compared to those for the compact-sphere clouds (Section \ref{sec:vertical}).
    %vertical structure of the fluffy-aggregate cloud is characterized by the large particle size and vertical extent (Figure \ref{fig:vertical}). 
    %Because of the low sedimentation velocity of aggregates, the porosity evolution leads to highly vertically extended cloud as compared to the classically assumed compact-sphere clouds.
%(3)  This facilitates the formation of high-altitude clouds.
    
(3) The fluffy-aggregate clouds largely obscure the absorption signatures of gas molecules in transmission spectra because of the large vertical extent if the aggregates are constituted by submicron monomers (Section \ref{sec:spectrum}).
Although the spectra in visible to near-infrared tend to be featureless, the fluffy-aggregate clouds become optically thin at longer wavelength ($\ga2~{\rm \mu m}$).
Future observations probing mid-infrared wavelength, such as JWST and ARIEL, may be able to detect molecular signatures even if the spectrum looks featureless in visible to near-infrared wavelength.

(4) CPAs also produce the spectral slope originated by the scattering properties of aggregates (Section \ref{sec:kappa_agg}). The slope reflects the wavelength dependence of the aggregate scattering opacity, $\alpha_{\rm c}\propto\lambda^{-2}$ (Section \ref{sec:spectrum}).
This could be potentially used as an observable signature of CPAs if the atmospheric scale height is well constrained.
    
(5) Comparing our synthetic spectra with the observations of GJ1214b, we find that the models of the high-metallicity atmospheres ($\geq100\times$ solar) well matches the observations if the CPAs are constituted by submicron monomers (Section \ref{sec:GJ1214b}).
This is due to the fact that the spectral slope produced by CPAs mismatches the observed flat spectrum as long as the atmospheric scale height is large.
The predicted high-metallicity atmosphere potentially suggests the presence of mechanisms regulating the gas accretion onto past GJ1214b.

We note that our results do not rule out other scenarios explaining the flat spectrum of GJ1214b, such as photochemical hazes \citep{Morley+15,Kawashima&Ikoma18,Kawashima&Ikoma19,Kawashima+19,Adams+19,Lavvas+19}.
The spectrum with hazes could also match the observations if the haze production rate is sufficiently high \citep{Lavvas+19}.
On the other hand, the hazes tend to produce the spectral slope caused by the haze opacity in the Rayleigh regime \citep{Kawashima&Ikoma18,Kawashima&Ikoma19,Lavvas+19}.
Therefore, from the same reason what we discussed for fluffy-aggregate clouds, the high-metallicity atmosphere may be still needed to explain the flat spectrum with photochemical hazes.
However, if hazes grow into moderately compressed aggregates ($D_{\rm f}\approx 2.4$), it could be possible to explain the flat spectrum with solar-metallicity atmosphere \citep{Adams+19}.
Exploring the porosity evolution of hazes will be helpful to constrain the atmospheric metallicity of GJ1214b.

The compact-sphere cloud is still not ruled out \citep{Gao&Benneke18}.
$K_{\rm z}$ for settling aerosols is still uncertain, and there is an order of magnitude uncertainty among different model predictions \citep[e.g.,][]{Komecek+19}.
Although GJ1214b is a close-in planet (semi-major axis is $0.014~{\rm AU}$), it might retain a non-zero eccentricity \citep{Charbonneau+09,Carter+11} that yields distinct atmospheric circulations \citep[e.g.,][]{Kataria+13,Lewis+17,Ohno&Zhang19} and possibly $K_{\rm z}$.
If it is possible to distinguish the fluffy-aggregate and compact-sphere clouds from observations, it might help to understand the aerosol transport processes in exoplanetary atmospheres.

\acknowledgments 
We thank Yasunori Hori for motivating this study and Yui Kawashima for helpful comments on the modeling of transmission spectra.
We also thank Ryan Macdonald, Graham Lee, and Xi Zhang for insightful comments.
{We are grateful to the reviewer, Peter Gao, for useful comments that greatly improved the paper.}
This work was supported by JSPS KAKENHI Grant Numbers JP18J14557 and JP18H05438.

\software{TEA \citep{Blecic+16}, TIPS \citep{Gamache+17}, Matplotlib \citep{matplotlib}}
%%%%%%%%%%%%%%%%%%%%%%%%%%%%%%%%%%

\appendix
\section{Aggregate Opacity for Other Mineral Clouds}\label{appendix1}
\begin{figure*}[t]
\centering
\includegraphics[clip,width=0.5\hsize]{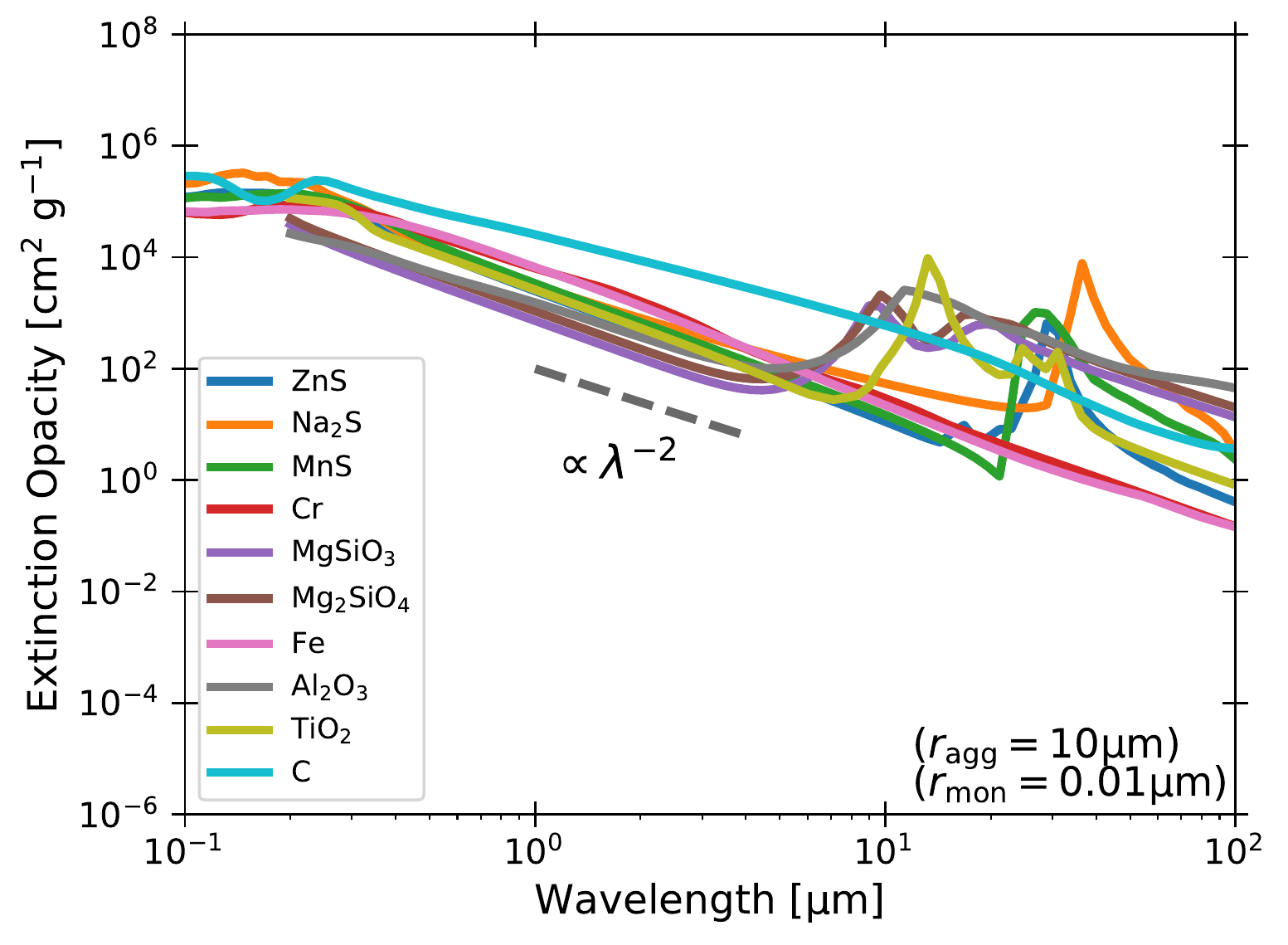}
\caption{Extinction mass opacity of aggregates with $D_{\rm f}=2$ for a variety of condensable materials.}\label{fig:appendix1}
\end{figure*}
{ 
In this appendix, we show the opacity of an aggregate made of various materials that potentially build up exoplanetary mineral clouds.
We have selected a variety of condensable materials (ZnS, $\rm {Na}_2S$, MnS, Cr, $\rm MgSiO_3$, $\rm {Mg}_2SiO_4$, Fe, $\rm {Al}_2O_3$) listed in \citet{Morley+12} and some nucleating species \citep[$\rm TiO_2$, C, see e.g.,][]{Helling+17,Helling+19}.
{The refractive indices of the  materials are} taken from \citet{Kitzmann&Heng18}.
Figure \ref{fig:appendix1} summarizes the calculated extinction opacities for $r_{\rm mon}=0.01~{\rm \mu m}$ and $r_{\rm agg}=10~{\rm \mu m}$. 
Some materials exhibit characteristic absorption {features} at $\lambda > 5~{\rm \mu m}$; for example, $\lambda\approx{40}~{\rm \mu m}$ for $\rm {Na}_2S$, $\lambda\approx{30}~{\rm \mu m}$ for $\rm MnS$, and $\lambda\approx{10}~{\rm \mu m}$ for $\rm MgSiO_3$.
Absorption also dominates over extinction at $\lambda<0.3~{\rm \mu m}$ for all materials.
On the other hand, the extinction opacity {at $\lambda=0.3$--$5~{\rm \mu m}$ is mostly dominated by}  scattering.
Therefore, many minerals other than KCl also produce an aggregate scattering slope of $\propto \lambda^{-2}$ (Section \ref{sec:kappa_agg}) at visible to near-infrared wavelengths. 
The exception we found is graphite, C, whose opacity is dominated by absorption even at near-infrared wavelength{s}.
}
%%%%%%%%%%%%%%%%%%%%%%%%%%%%%%%%%%%%%%%
%\bibliography{reference}

\end{document}